\documentclass{jfm}
\usepackage{graphicx}
\usepackage{gensymb,amsmath,amssymb,hyperref}
\usepackage{color}
\usepackage{blindtext}
\usepackage{lineno}
\usepackage{cancel}
\usepackage{soul}
\usepackage[normalem]{ulem}
\usepackage{pdfpages}

\definecolor{darkblue}{rgb}{0.2,0.2,0.6}

\hypersetup{colorlinks,breaklinks,
            linkcolor=darkblue,urlcolor=darkblue,
            anchorcolor=darkblue,citecolor=darkblue}

\usepackage{epstopdf, epsfig}
\usepackage{xcolor}
\usepackage{comment}

\shorttitle{Wedge \& cone slamming on water}
\shortauthor{U. Jain et al.}

\title{On wedge-slamming pressures}

\author{Utkarsh Jain\aff{1}
  \corresp{\email{u.jain@utwente.nl}},
  Vladimir Novakovi\'{c}\aff{2}, Hannes Bogaert\aff{2}
 \and Devaraj van der Meer\aff{1}}

\affiliation{\aff{1}Physics of Fluids Group and Max Planck Center Twente for Complex Fluid Dynamics, MESA+ Institute and J. M. Burgers Centre for Fluid Dynamics, University of Twente, P.O. Box 217, 7500AE Enschede, The Netherlands 
\aff{2}Maritime Research Institute (MARIN), 6708PM Wageningen, The Netherlands}

\begin{document}

\maketitle

\begin{abstract}
The water entry of a wedge has become a model test in marine and naval engineering research. Wagner theory, originating in 1932, predicts impact pressures, and accounts for contributions to the total pressure arising from various flow domains in the vicinity of the wetting region on the wedge. Here we study the slamming of a wedge and a cone at a constant, well-controlled velocity throughout the impact event {using high fidelity sensors}. Pressures at two locations on the impactor are measured during and after impact. Pressure time series from the two impactors are discussed using {inertial} pressure and time scales. The non-dimensionalised pressure time series are compared to {sensor-integrated} {averaged} composite Wagner solutions \citep{zhao_faltinsen_1993}, {\citet[4.7]{logvinovich1969hydrodynamics}, modified Logvinovich \citep{korobkin2005modified} and generalised Wagner models \citep{korobkin_2004}. In addition, we provide an {independent} experimental justification {of approximations made in the literature in extending the Wagner model to three-dimensions.} {The second part of the paper deals with pre-impact air cushioning - an important concern since it is responsible for determining the thickness of air layer trapped upon impact. Using a {custom-made} technique we measure the {air-water interface} dynamics as it responds to the build up of pressure in the air layer intervening in between the impactor and the free surface. We show both experimentally and using two-fluid boundary integral (BI) simulations, that the pre-impact deflection of {the} interface due to air-cushioning is fully described by potential flow.}}

\end{abstract}


\section{Introduction}

Model tests for measuring the hydrodynamic loading on a solid structure impacting on water are relevant for situations such as in naval engineering \citep{kapsenberg2011slamming, abrate2011hull}, offshore \& ocean engineering \citep{faltinsen2000hydroelastic}, sloshing in liquid cargo containments \citep{dias2018slamming}, and in aerospace engineering \citep{SEDDON20061045}. From a large catalogue of simple shapes of the impactor that can be chosen, impact tests with a wedge have become {the standard} in the industry. The first pioneering work to model impact force on the wedge in the context of planes landing on water was done by \citet{von1929impact}. The main contribution to loading was understood to be an added-mass effect due to the liquid being displaced by the submerging wedge. However, as the wedge enters water, the displaced liquid piles up at the edge of the wetted region on the wedge's surface. As a result, the rising liquid wets a larger portion of the surface of the wedge than the extent of submergence of the wedge would imply. Wagner's treatment \citep{wagner1932stoss} of the same problem (a wedge with a small deadrise angle) provided a significant improvement on the previous approach by accounting for this liquid pile-up. {The resulting pressure distribution from the treatment is singular at the point of the liquid-solid intersection; this is indicative of how much energy is transferred by the rising liquid to the body {near} the intersection point ({trailing edge of the jet}), thus highlighting the importance of modelling this region in detail.}

The wedge problem was attractive for modelling as the flows resulting from impact are essentially two-dimensional. Wagner's potential flow approach was particularly successful as the model could be further extended to three dimensions \citep{shiffman1951force, scolan_korobkin_2001}, for better modelling of the finer aspects of different flow domains in the liquid, and improved predictions of the force and pressure distribution {\citep[4.7]{logvinovich1969hydrodynamics}, \citep{dobrovol'skaya_1969, pierson1950penetration, korobkin_2004}}. Several numerical tools have been developed dedicated to resolving such fine details of the flow domain that play an important role in the early stages of loading \citep{WU2004277,faltinsen2005generalized, PESEUX2005277, WANG2015181}. However, a limited number of experimental works validate the analytical models, and moreover, fewer even provide a direct comparison of predicted and measured peak pressures{, let alone a full comparison of the measured and theoretical pressure time series}. {The present work fills this crucial gap by performing {high temporal resolution, well-controlled, constant velocity} impact experiments and providing comparisons of measured {pressure time series} with the {original} \citet[4.7]{logvinovich1969hydrodynamics}, {modified Logvinovich} \citep{korobkin2005modified}, generalised Wagner \citep{korobkin_2004} and asymptotically matched {composite} \cite{zhao_faltinsen_1993} models.}

\citet{chuang1966deadrise, chuang1966experiments} and \citet{chuang1971cones} performed several insightful experimental studies with different geometries and deadrise angles of the impactor. Works by \citet{TVEITNES20081463} and \citet{vincent_xiao_yohann_jung_kanso_2018} focussed on the total impact force on the wedge. Experiments by \citet{takemoto1984some} validated the Wagner profile (pressure distribution along the wetted surface of the wedge). \citet{yamamoto1984water} also performed a large number of experiments, cataloguing peak pressures at different locations on a wedge, and how they change with deadrise angle and impact velocity. \citet{donguy2001numerical}, \citet{PESEUX2005277} and \citet{debacker2009experimental} performed experiments with axisymmetric impactors and provided comparisons with numerical models. \citet{yettou2006experimental} and \citet{lewis2010impact} provided detailed information on the variation of the impact pressures along a wedge's surface. {\citet{GREENHOW1987214} performed insightful studies investigating the pressure distribution on a water impacting wedge and comparing them to exact solutions by \citet{dobrovol'skaya_1969}.}

Here we perform water-impact experiments with a wedge {of} deadrise angle of 10$\degree${, and cones of deadrise angles 5, 10, 20 and 30$\degree$}. The impact velocities are controlled using a linear motor such that they are maintained at a constant value during impact with a relatively high degree of precision (see appendix \ref{sec:wedgesappendix}). Pressures at two locations each, on either side of the wedge are measured using high temporal resolution pressure sensors. We show using non-dimensional scales that a sudden under-pressure prior to the impact peak, reported by several prior studies \citep{lewis2010impact,debacker2009experimental,PESEUX2005277,donguy2001numerical,tenzer2015experimental}, {is systematic, scales as $\rho V^2$, and likely has a hydrodynamic origin}. {Maximum pressure coefficients} {from {the} original \citet[4.7]{logvinovich1969hydrodynamics}, modified Logvinovich \citep{korobkin2005modified}, {and} generalised Wagner models \citep{korobkin_2004}, {an exact self-similar integral formulation by} \citet{dobrovol'skaya_1969}, asymptotic solutions from \citet{WANG201723}, and the composite solution of \citet{zhao_faltinsen_1993} are compared.} {However since the measured pressures are naturally {integrated over the sensor area and} averaged, the above models are also used to compute space-averaged results, {which} are directly compared with the measured time series. Such space-averaged computations from modified Logvinovich model were also discussed by \citet{scolan2012low}, and actually compared against experiments by \citet{SCOLAN2014470,YVESMARIE20141197}.}

{Owing to {high accuracy, highly time-resolved} experiments, we are able to comment on pressure features such as the pre-peak under-pressure, and the variation of the pressure along the width of the body with some confidence, something that is not available in existing literature. {For impacting} cones we perform a similar comparison of measured pressure {time series} and space-averaged results {against the} composite solution of \citet{zhao_faltinsen_1993}.} Finally, by {directly} comparing the {measured} peak pressures from wedge and cone, we {directly and independently verify} the approximation in {which the} three-dimensional {jet} {flow is treated} as quasi-two-dimensional. 

{The final part of the paper deals with air cushioning prior to the moment of impact. Air entrapment under the impactor can completely alter the liquid surface's shape, and thus, pressures at impact. Entrapment may occur depending on how air cushioning prior to impact creates a depression on the free surface.} {While} no air-trapping effects upon impact are anticipated {for the wedges and cones with large deadrise angles, we show using an in-house measuring technique that the air cushioning effect under the impactor results in downwards deflection of the water surface that is of the order of $10^{-3}$ times the cone diameter. As far as the interface's deflection due to air-cushioning is concerned, stagnation point flow at the interface, and hence potential flow, fully describes the process. We support our measurements with two-fluid boundary integral simulations{, where both the air and water phase flows are treated as potential flows}.} Although the extent of this depression {of the free surface} may seem insignificant at first sight, it should be stressed that it occurs at the site of the first contact, where the pressures occurring are expected to be largest. {In fact pressure time-traces obtained by \citet{nethercote1986some} at the wedge keel were indeed larger than impact pressures along the wedge's sides. Interestingly they observed some oscillations in pressures at the centre, which were inconclusively suggested to result from air entrapment at the keel. Our experiments were done with the same range of deadrise angles as \citet{nethercote1986some}, and we find no visible air entrapment for deadrise angles larger than $1\degree$ ({as discussed in more detail in the} supplementary text).} {Since the {air cushion} already sets the target surface into motion before the moment of impact, air cushioning will likely} have a mitigating effect on the pressure {maximum} at the keel. {{Finally,} force measurements on the cones are also performed, and briefly discussed in the appendix where they are also compared to the \citet[4.9]{logvinovich1969hydrodynamics} model {and the composite solution of \citet{zhao_faltinsen_1993}.}}

\section{Pressures at discrete locations}

\subsection{Setup description}\label{sec:conewedgesetup}

\begin{figure}
    \centering
    \includegraphics[width=.9\linewidth]{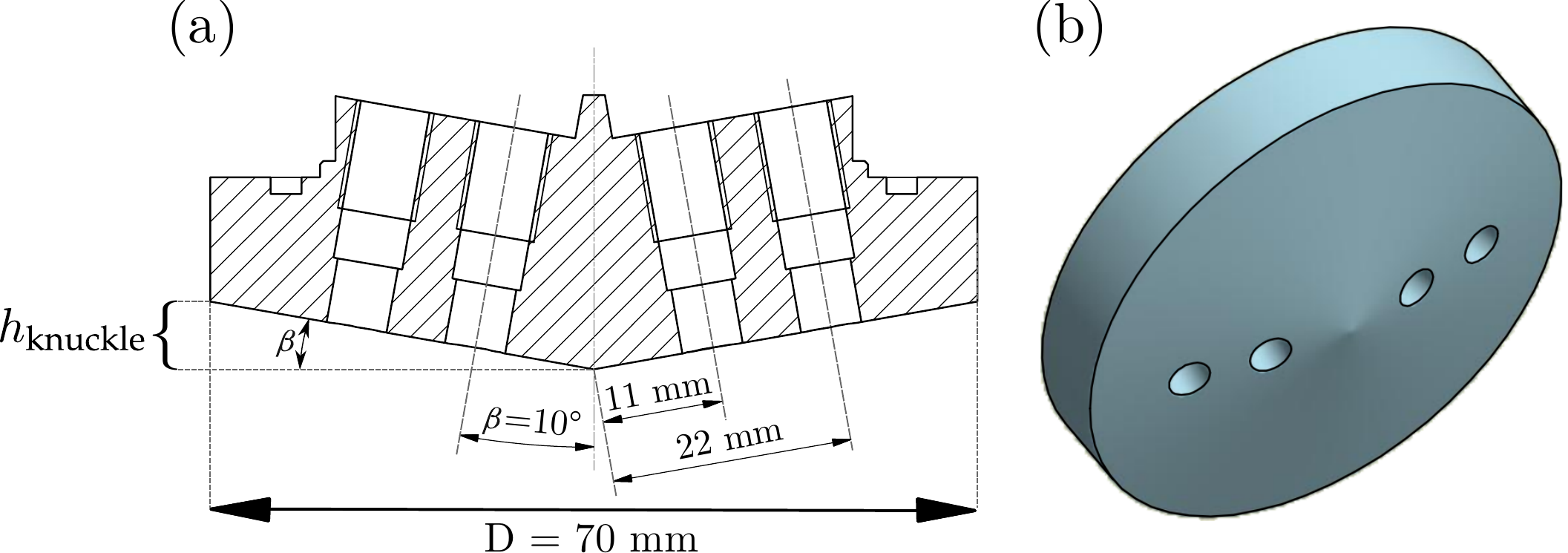}
    \caption{(a) Dimensions of the cone and wedge that are used for experiments. Pressure transducers are installed such that they are flush mounted with the surface on the impacting side. The impactor is closed on top such that the installed sensors installed are protected. (b) The cone's design is shown for illustration. Its bottom tip is called the keel. The location where its sloping surface sharply turns away into the vertical cylindrical surface is known as its knuckle.}
    \label{fig:conedesign}
\end{figure}

The cone used for experiments is 70 mm wide, and has a deadrise angle of 10$\degree$. The wedge used in experiments has the same deadrise angle, an impacting cross-section that is 70 mm wide, and a length of 140 mm. The overall dimensions are shown in figure \ref{fig:conedesign}(a). Both the wedge and cone house Kistler 601CAA dynamic piezoelectric transducers that are flush mounted with the impacting bottom surface. Vaseline is used to seal any gaps between the sensors and the mounting sites on the impactors. All the sensors had a sensing area 5.5 mm in diameter, and a peak acquisition rate of 200 kHz. Two sensors each are installed on either side of the keel such that the pressures measured on one side of the cone/wedge can be corroborated by measurements along the opposite side, a posteriori verifying the horizontality of impact. The first two sensors are installed such that the centers of their sensing area lie at a distance of 11 mm from the keel. The second set of sensors are installed a further 11 mm from the first pair of sensors. An illustration of the cone is shown in figure \ref{fig:conedesign}(b). A wedge is simply a 2D projection of the cone: sensors on the wedge are installed at the same distance from the keel as in the cone. Note that, in addition to these flush-mounted transducers at the impacting surface, additional `placebo' sensors are installed in both impactors to check that the pressures \emph{at impact} are not affected by the impactor's change in acceleration. The reader is referred to appendix \ref{sec:wedgesappendix} for discussion. The impactors are designed such that they are closed from the top to protect the sensors. Both were fabricated out of PET polyester to minimise the weight of the moving parts. This is done so as to maximise the velocity with which they can be translated. {The mass of the wedge assembly was 795 g, and that of cone was 611 g.}

The impactor is mounted on a linear motor using an aluminium rod. The assembly’s position can be controlled with an accuracy of approximately 0.6 $\mu$m. {The degree of good velocity control is indicated by showing how repeatable the peak pressures are at a resolution of 200 kHz in figure \ref{fig:conepressures}}. The acceleration that is generated by the linear motor in the course of its stroke is limited by the current it can draw. Thus the linear motor’s motion is programmed such that it achieves a specified velocity when in the middle of its available stroke. Concerning the wedge (or cone), the stroke is programmed such that it attains a constant velocity $V$ before it reaches the water free surface, and maintains so while the impactor plunges into the target pool until its submergence. $V$ is varied from 0.6 -- 4.5 m/s. The target liquid bath consisted of de-mineralised water in a glass tank of area 50 cm $\times$ 50 cm, and a depth of 30 cm.

\begin{figure}
    \centering
    \includegraphics[width=.99\linewidth]{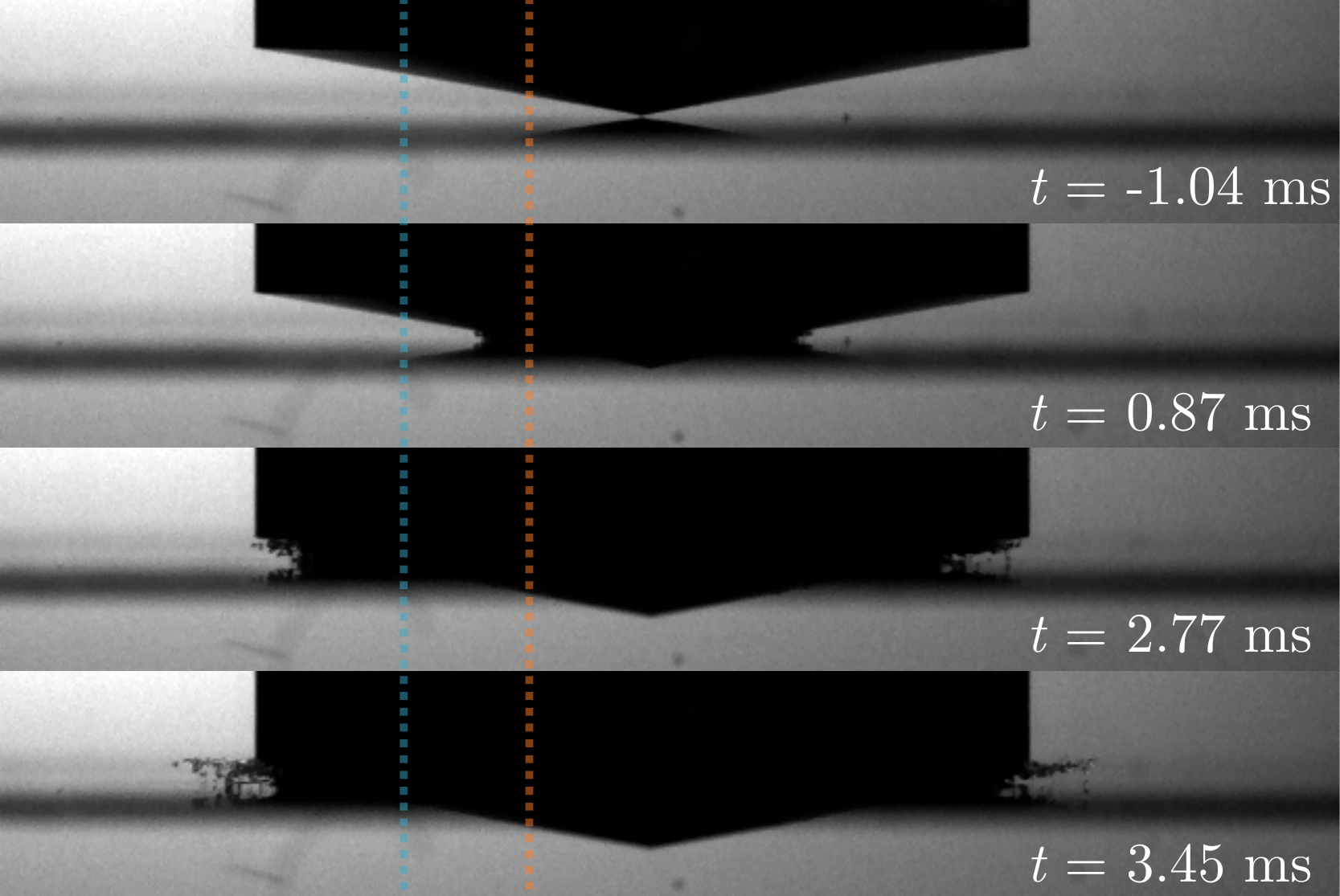}
    \caption{Snapshots showing the early stages of a cone entering water at 1.0 m/s. {The time series are centred around} at $t=0$ ms, {which is defined as the moment when the} {pressure on the first sensor from the keel starts to rise}. The orange and blue dotted lines show the locations of the pressure sensors whose centers are positioned 11 mm and 22 mm from the keel respectively. At $t =$ {0.87} ms the cone has travelled a distance at which a stationary water surface should reach the center of the diaphragm of the first sensor. At $t = $ {2.77} ms, a stationary water surface would reach the center of the second sensor's sensing area. At $t= $ {3.45} ms, the rising water is seen to separate from the cone. Without the local pile-up of water, the stage seen at $t = $ {3.45} ms, when the water separates from the cone's knuckle, would occur {5.13} ms after the start of cone's water entry.}
    \label{fig:conesnapshots}
\end{figure}

\subsection{Typical pressure time series {and repeatability}}
\subsubsection{Cones}

\begin{figure}
    \centering
    \includegraphics[width=.99\linewidth]{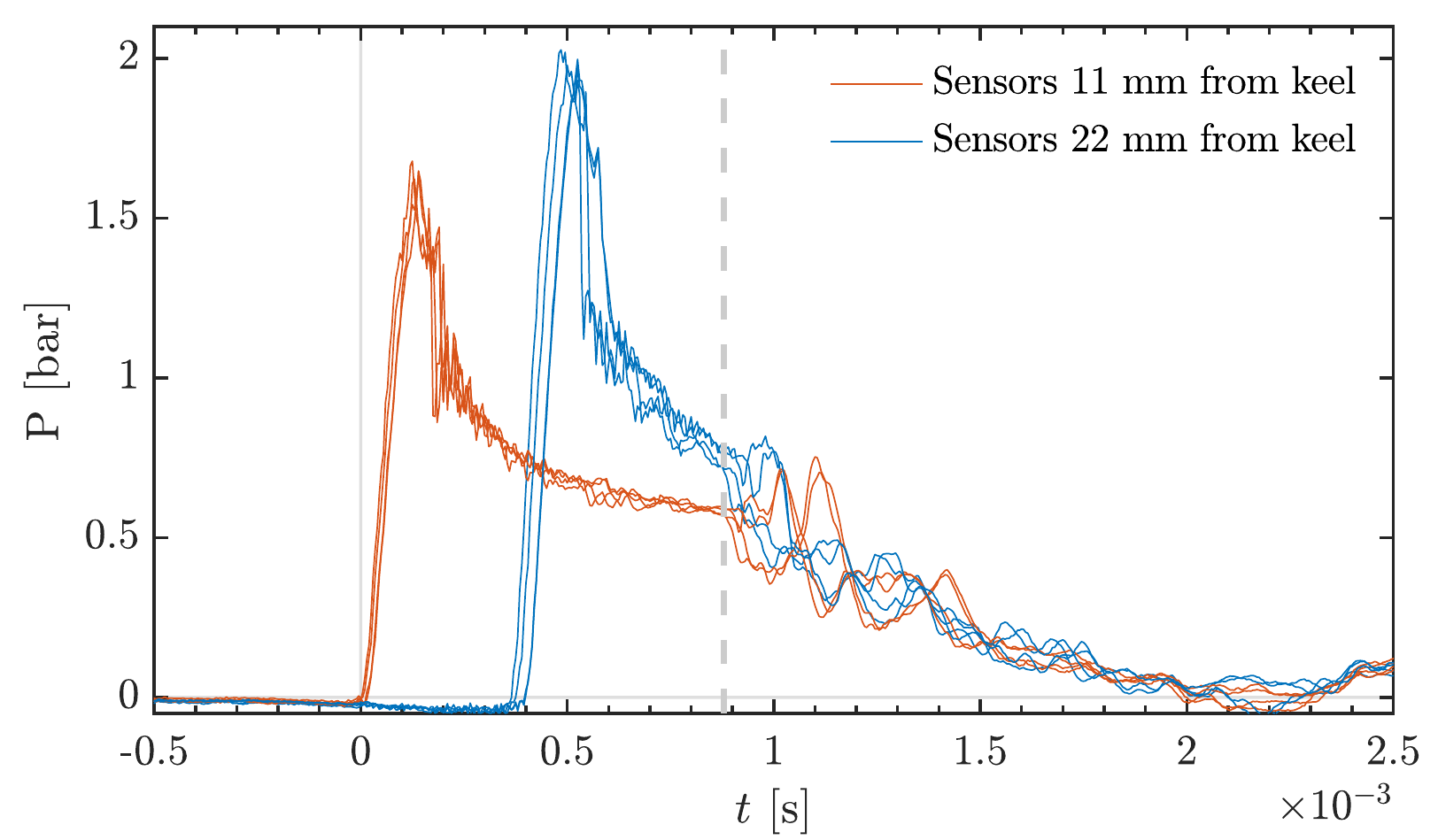}
    \caption{Pressures time series recorded with the cone impacting a water surface at a controlled velocity of 4 m/s. The data are shifted in time such that at $t =$ 0 ms, the water reaches the sensor(s) close to the keel (11 mm away along the cone contour), and the pressure here starts to rise. Approximately 0.34 ms later, the water reaches the second sensor, causing the pressure to rise there. Notice that immediately before the pressure starts to rise on the second sensors, it becomes negative. This transient under-pressure is created due to jet flow. A dashed-line at $t = $ 0.88 ms marks the time when the pressure reading on all the sensors starts to drop suddenly and simultaneously. This corresponds to the stage shown in the last panel of figure \ref{fig:conesnapshots} when the rising water reaches the end of the contour, and separates at the cone's knuckles. {Four realisations of the measurements are shown as an indicator of repeatability. Note that within each peak the agreement could even be further improved by performing a small time-shift.}}
    \label{fig:conepressures}
\end{figure}

We denote the events that occur between the first touch-down of the cone's keel, until the time when flow separates from its knuckles, as events during impact. This stage in experiments was recorded from a side view at 40,000 frames per second. Select key stages from an experiment at $V = $ 1 m/s are shown in figure \ref{fig:conesnapshots}. The two dotted lines in the figure mark the radial locations of sensors {present} on {either} side of keel. {During a time interval of} 1.91 ms from the start of the cone's entry into water {at ($t = 0.87$ ms in figure \ref{fig:conesnapshots}}), it moves a vertical distance of 1.91 mm into the liquid bulk, such that the cone ought to be submerged until the center of the first sensor (location highlighted by orange dotted line). However it can be seen from figure \ref{fig:conesnapshots} that at this stage the rising liquid has already risen past the marked location. A similar observation can be made from the succeeding snapshot at $t = 2.77$ ms {(3.82 ms from the start of cone's entry)}: when $V  t$ would suggest the water level to have reached the blue dotted line, the liquid has risen well past the center of {the} second sensor. The reason for this observation is as follows: as the cone enters into water, it displaces the water. Due to the liquid being impulsively displaced, it rises along the bottom surface, and emerges as a jet. This rising jet of liquid causes the cone's bottom surface to wet faster than the rate at which it would wet if the cone were to submerge quasi-statically. The liquid rising along the cone turns away from its contour, feeding mass to the emerging jet. This region of the liquid domain is known as jet root region (see figure \ref{fig:wagnersketch}). The maximum impact pressure {at a point occurs} when it falls in the jet root region \citep{wagner1932stoss}. Typical pressure measurements are shown in figure \ref{fig:conepressures}.

The final stage is reached when the rising liquid reaches the cone's knuckle, and separates from the cone's contour. As shown in the last panel of figure \ref{fig:conesnapshots}, this stage is reached 3.45 ms after the start of cone's entry into water, while $V  t$ would suggest this to occur at $\Delta t = h_{\text{knuckle}}/V =6.17 $ ms after impact, i.e., at $t =$ 5.13 ms.

{T}he pressure time series are shifted to $t=0$ at the point when the pressure on the first sensor from the keel starts to rise. A typical set of measurements from one experiment are shown in figure \ref{fig:conepressures}. A large $V$ example is shown in figure \ref{fig:conepressures}.

In figure \ref{fig:conepressures}, the respective data from sensor(s) are shown using the same colour-coding as their location highlights in figure \ref{fig:conesnapshots}. Before the readings on sensors at either location start to rise, the pressure becomes slightly negative. This feature is more prominently visible in the data from the second sensor(s) from the keel. This has been ascribed to jet flow of the liquid \citep{PESEUX2005277}. The jet precedes the oncoming thicker bulk of liquid which constitutes the `inner-region' (see figure \ref{fig:wagnersketch}). This pressure drop is regularly observed in experimental studies of wedge and cone impacts on water \citep{debacker2009experimental, tenzer2015experimental, PESEUX2005277, donguy2001numerical, lewis2010impact}. However, to the best of the {authors'} knowledge, its origin has not been systematically identified. \citet{lewis2010impact} in particular suggested several sources for this pressure drop - from erring instruments to aerodynamic effects due to a `high-speed flow over the pressure sensor'. {In section \ref{sec:wedgepressurecomps} we discuss the same finding. We are able to show that these pre-peak under-pressures in fact scale neatly with the dynamic pressure scale, but we are unable to conclusively understand {their} origin}.

In the next stage, the pressure rises to its peak value, denoted as $P_{\text{peak}}$. Naturally, the sensors farther from the cone's keel register the peak at a later time than the first pair of sensors. After the peak, the pressure decreases towards a plateau value corresponding to the typical hydrodynamic pressure experienced by the sensors while the cone plunges further into the water bath. The final stage is reached when the readings on all the sensors near-simultaneously drop down from the value to which they appeared to be saturating. This point is highlighted in figure \ref{fig:conepressures} using the grey dashed line, and corresponds precisely to the stage when the rising liquid jet reaches the cone's knuckle and separates from it. \citet{PESEUX2005277} made the same observation regarding the sudden pressure drop. The small temporal inconsistencies that can be observed in measurements from different sensors at the same radial location can be attributed to an unavoidable, but small asymmetry in the angle with which the cone impacts.

\subsubsection{Wedge impact pressures}

\begin{figure}
    \centering
    \includegraphics[width=.99\linewidth]{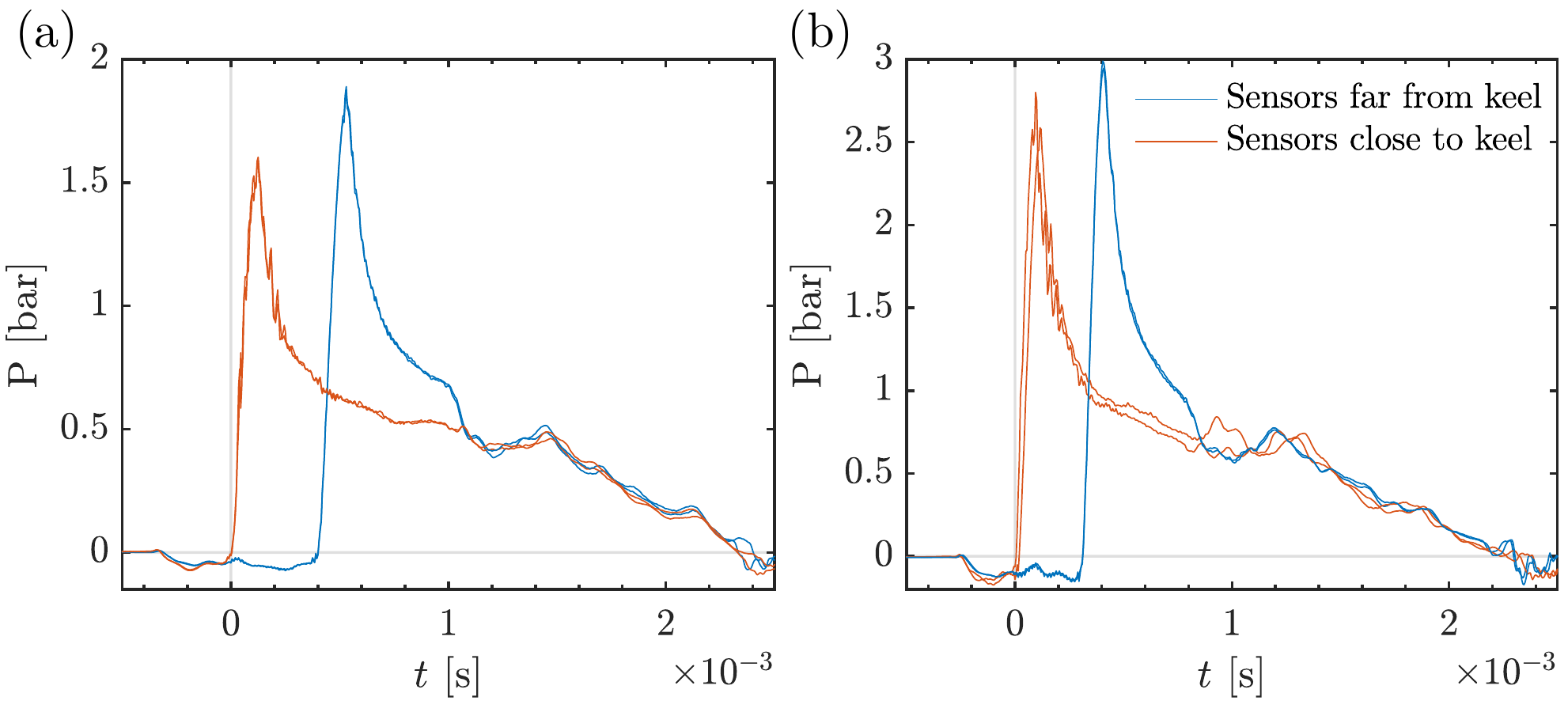}
    \caption{Pressures time series recorded with the wedge impacting a water surface at controlled speeds of (a) 3.0 m/s and (b) 4.0 m/s. The data are shifted in time such that at $t =$ 0 ms, the water reaches the sensor(s) close to the keel (11 mm away along the cone contour), and the pressure here starts to rise. Notice that in both the plots, the pressures suddenly drop below zero for a short time before impact due to jet flow. The effect is substantially larger than in the case of the cone, owing to the wedge impact being a (quasi) two-dimensional process. In fact, the point at which the pressures suddenly drop is the time when the wedge starts to enter water (also see inset in figure \ref{fig:wedgewagner}(a)). The legend is shared between the two panels.}
    \label{fig:wedgepressures}
\end{figure}

Experiments with impacting wedges on water are done with the impactor as described in section \ref{sec:conewedgesetup}. The flows caused by a wedge impacting on water can be approximated as a two-dimensional phenomenon, and be more readily described by an inviscid flow model. As {with} the experiments described in {the} previous section with a cone, we measure pressures along the wedge's impacting surface at two locations. Typical measurements are shown in figure \ref{fig:wedgepressures}. As with measurements from the cone (figure \ref{fig:conepressures}), the pressure readings suddenly become negative immediately prior to rising again, which again, is ascribed to the initial jet flow. Notice that compared to the similar under-pressure seen for cone impact from figure \ref{fig:conepressures}, the measurements from the wedge show both a more abrupt, and a larger drop in pressure. This suggests that the jet and other flows created by the impact of a wedge are more violent than those due to a cone. Indeed it can be shown analytically that for a given deadrise angle, and at the same point in time, the rise of water is higher for a wedge than for a cone \citep{Schmieden1953, faltinsen1998water}. From our measurements one can also notice that the peak pressures measured on a wedge are higher than those attained with a cone at the same radial locations, which we will discuss in more detail in section \ref{sec:3dfloweffect}. It can be seen from both the experiments shown in figure \ref{fig:wedgepressures} that the pressure-drop prior to its rise occurs at nearly the same time at different distances from the wedge's keel. Compared to this, there is a significant time delay between when the impact pressure peaks are registered on sensors at varying distances. This is an indication that the initial liquid jet {(which, according to \citet{PESEUX2005277} and \citet{lewis2010impact} causes the under-pressure)}, emerges much faster than the later rise of the root of the jet that causes the peak impact pressure.

The successive stages show the pressure saturating to a steady value for a short time. The final stage is observed when the pressures suddenly drop (at approximately $t= 1.3$ ms in figure \ref{fig:wedgepressures}(a) and $1.2$ ms in figure \ref{fig:wedgepressures}(b)) when the rising liquid reaches the wedge's knuckle and separates from its surface. Thus it can be seen that until after the time of attainment of peak impact pressure, the contributions to pressure at a point on the wedge arise from different hydrodynamic sources. The model by \citet{wagner1932stoss} identifies and models the different contributions to pressure at a point on a wedge. This is discussed in the following section.

\subsection{Comparisons of pressures from different models}\label{sec:wedgepressurecomps}

\citet{wagner1932stoss} derives the impact pressures on a water-entering object due to resulting flows from first principles in the case of an ideal fluid. It has proven to be a particularly successful model to which {numerous extensions and} improvements could be made (e.g.,~ \citet{armand1986hydrodynamic,howison_ockendon_wilson_1991,scolan_korobkin_2001,korobkin_scolan_2006,iafrati2008hydrodynamic}). Here we briefly {mention} the main results from the derivation of the pressure distribution on a water-entering wedge.

\begin{figure}
    \centering
    \includegraphics[width=.70\linewidth]{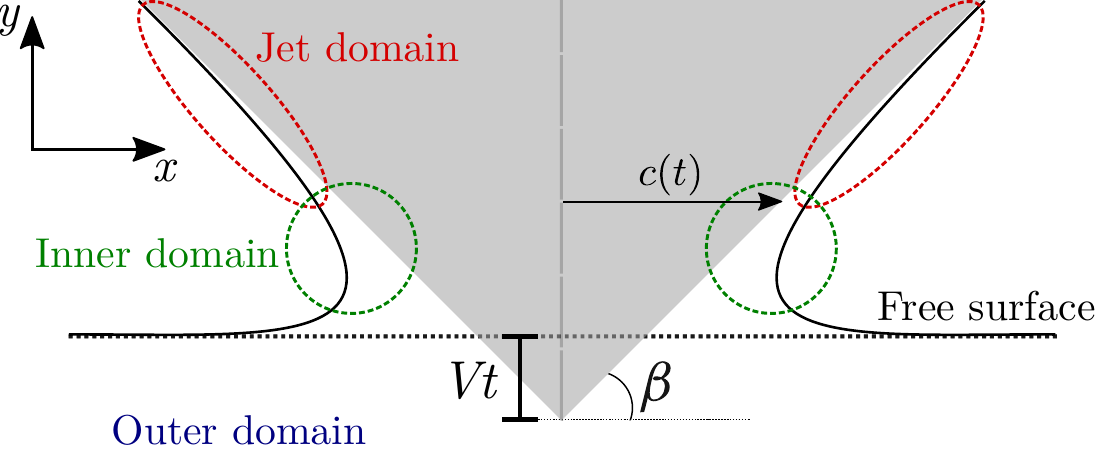}
    \caption{{Schematic and} definitions of the different flow regions as used in Wagner model for a wedge impacting in water. The inner domain is also variedly known as jet root region.}
    \label{fig:wagnersketch}
\end{figure}

\begin{figure}
    \centering
    \includegraphics[width=.99\linewidth]{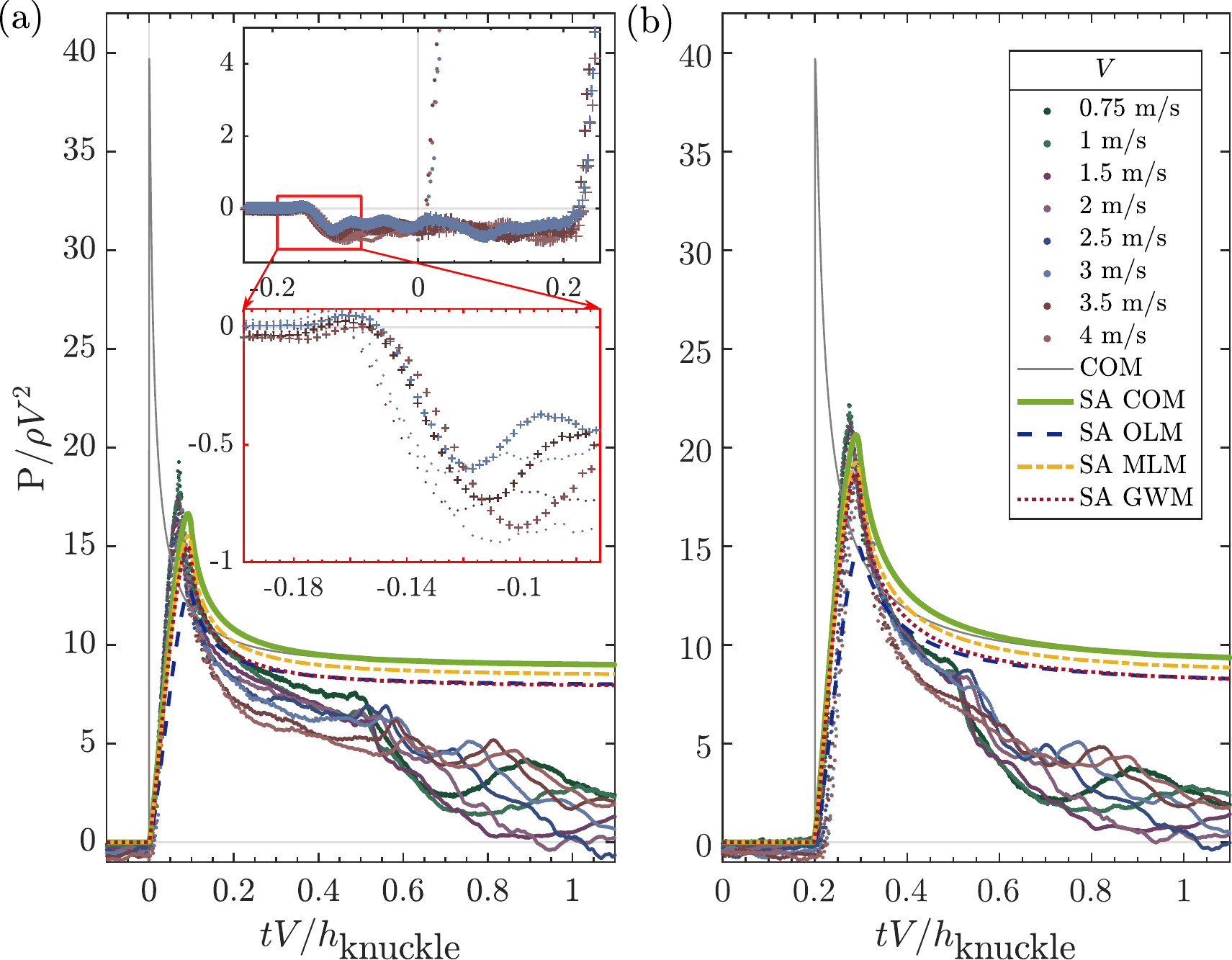}
    \caption{For the wedge case, pressure measurements of the type shown in figure \ref{fig:wedgepressures} {for} a range of {impact velocities} $V$ are non-dimensionalised and plotted from the first sensor in panel (a), and from the second sensor in panel (b). As in earlier plots, the {experimental} time series were shifted so that the point at which the reading on the first sensor starts to rise lies at $t = $ 0 ms. 
The insets in panel (a) shows a zoomed-in region of the data shown in both the panels in the vicinity of $t  V / h_{\text{knuckle}}=0$. {Dots and plusses in the inset represent the data from first and second sensors respectively.} The data that start to rise at $t  V / h_{\text{knuckle}} = $ 0 are the readings from the sensor close to keel, the remainder at the data from the other sensor. The quantity $h_{\text{knuckle}} = D \tan \beta /2 \approx$ 6.17 mm. The grey solid curves are {the point-pressure} computations from the composite solution (equation \eqref{eqn:compositepressure1}{, also see row 5 and column 7 in table \ref{tab:cpmaxcoeffs}}). {The final four curves are the space-averaged pressures from {the composite solution (SA COM,} green line, \eqref{eqn:compositepressure1}), {the original Logvinovich model (SA OLM,} blue-dashed, \eqref{eqn:OLM}), {the modified Logvinovich model (SA MLM,} yellow dashed line \eqref{eqn:MLM}) and {the generalized Wagner model (SA GWM,} red dotted line, \eqref{eqn:GWM}). {Note that, of course, these theoretical time series are shifted in time by the same amount as the experimental ones.}}}
    \label{fig:wedgewagner}
\end{figure}

When the wedge enters water, a part of its surface is wetted. This wetted length of the wedge in the horizontal direction is treated as a plate. As the wedge further progresses in the liquid bulk, the wetted region, or length of the plate $2c(t)$, expands. A sketch of the situation is shown in figure \ref{fig:wagnersketch}. The resulting fluid flow problem posed in the model is one to calculate the flow potential due to an expanding plate at the fluid-surface. The fluid is assumed to be inviscid and incompressible such that the flow potential $\phi$ satisfies Laplace's equation \begin{equation}
    \nabla^2 \phi = 0.
\end{equation} The pressure everywhere in the fluid is given by unsteady Bernoulli equation \begin{equation}
    \frac{1}{\rho}p(x,y,t) = - \frac{\partial \phi}{\partial t} - \frac{1}{2} \lvert \mathbf{\nabla} \phi \rvert^2.
\end{equation}
The liquid interface directly under the plate (region $x<c$) obeys the kinematic condition $\left(\partial \phi / \partial y \right)_{y=0} = -V $. Along the free surface (region $x>c$) away from the plate, $\phi(x,0) = 0$. Without repeating the complete derivation \citep{wagner1932stoss}, $\phi$ may be written as the real part of the complex potential \begin{align}
    \mathcal{W}(z) = i V \left( z + \sqrt{z^2 - c^2} \right),
\end{align} 
where $z = x + iy$, and $c$ is the half-length of the wetted plate (shown in figure \ref{fig:wagnersketch}). The point $c$ lies at the jet root, approximately at the intersection of the inner and jet domains in figure \ref{fig:wagnersketch}. By computing \begin{align}
    \frac{d \mathcal{W}}{dz} = \frac{\partial \phi}{\partial x} - i \frac{\partial \phi}{\partial y} = u - i v
\end{align}
on the $x$ axis, we find for $x<c$ that \begin{align}
    \left( \frac{\partial \phi}{\partial y} \right)_{y=0} = - V,
\end{align}
and
\begin{equation}
    \frac{\partial \phi}{\partial x} (x,0) = V \frac{x}{\sqrt{c^2 - x^2}}.
\end{equation}
Noting that $\phi(x,0) = \text{Re}\left[ \mathcal{W}(x,0) \right] = - V \sqrt{c(t)^2 - x^2}$, the unsteady Bernoulli equation yields the pressure on the body as
\begin{align}
    {p} &= \rho \frac{\partial}{\partial t} \left( V \sqrt{c^2 - x^2} \right) -  \frac{\rho}{2} \left( V \frac{x}{\sqrt{c^2 - x^2}} \right)^2 \label{eqn:wagnerpressure1}\\ 
    &= \rho\frac{dV}{dt}\sqrt{c^2 - x^2} + \rho \frac{V c}{\sqrt{c^2 - x^2}}\frac{dc}{dt} -  \frac{\rho}{2} \frac{V^2 x^2}{c^2 - x^2}. \label{eqn:wagnerpressure2}
\end{align}
The first term above becomes zero for constant $V$. The second term is known as the slamming pressure, where the dependence on the wetting rate $dc/dt$ plays a crucial role. It predominantly originates from the outer domain and is associated with accelerating the added mass of liquid. {The final term is the jet pressure.} Wagner's {central} contribution was to determine the wetting rate $c(t)$ for certain geometries. For a wedge {with deadrise angle $\beta$}, it is found to be \citep{wagner1932stoss} \begin{equation}\label{eqn:wagnerconditionwedge}
    c(t) = \frac{\pi V t}{2 \tan \beta}.
\end{equation}
Dropping the first term from equation \eqref{eqn:wagnerpressure2} for constant $V$, the expression for total pressure becomes \begin{equation}\label{eqn:wedgetotalpressure}
    p(x,0) = \rho V^2 \left[ \frac{\pi}{2\tan \beta } \frac{c}{\sqrt{c^2 - x^2}} - \frac{1}{2} \frac{x^2}{c^2 - x^2} \right] \text{, for } (x<c).
\end{equation}
Note that this equation can be written as \begin{equation}
    \frac{p}{\rho V^2} = f(x/c),
\end{equation}
which suggests a dimensionless form introducing \begin{align}
   \Tilde{p} = \frac{p}{\rho V^2}, \; \Tilde{x} = \frac{x}{D}\; \text{, and } \Tilde{t} = \frac{Vt}{h_{\text{knuckle}}},
\end{align}
with which $x/c =  4 \Tilde{x}/\pi \Tilde{t}$. Accordingly, we non-dimensionalise pressure measurements from experiments done over a range of $V$ with the pressure scale $\rho V^2$, and plot them against time rescaled with $h_{\text{knuckle}}/V$ in figure \ref{fig:wedgewagner}. At each of the locations, the measurements are very convincingly collapsed by the re-scaling described above. {Note {that} the timeseries are shifted in time such that the impact on first sensor is shown to occur at $t=0$ {in figure \ref{fig:wedgewagner}, and the model calculations are shifted by the same amount.}}

{In {the two insets} of figure \ref{fig:wedgewagner}(a) we show that the under-pressure that occurs prior to the peak, {also collapses when rescaled using} inertial scales. Clearly the occurrence of this under-pressure is systematic, and the rescaling indicates a hydrodynamic origin{, possibly connected to the jet region moving over the pressure sensor. This is also consistent with the observation that the pressure drops slightly earlier in the first sensor and subsequently in the second}. {The {data} collapse {obtained} with the {inertial pressure and time scales} 
is consistent with {the theoretically expected self-similarity of the impact as a whole.}}
However since the expression \eqref{eqn:wedgetotalpressure} only concerns the region between the turnover points at the jet root, this under-pressure, although evidently also inertial, is not {included} 
in the above expression.}

{\subsection{The singularity at $x = c$, and comparisons of measured pressures with models}}

\subsubsection{Composite model from asymptotic matching by \citet{zhao_faltinsen_1993}}
\label{sec:COMP}

The singularity {occurring} in equation \eqref{eqn:wedgetotalpressure} {for $x \rightarrow c$} hinders one from computing a {peak impact pressure value} that can be directly compared to experiments. \citet{armand1986hydrodynamic}, \citet{wilson1989mathematics} and \citet{howison_ockendon_wilson_1991} described how to match the solutions in inner and outer domains using asymptotic expansions. The analysis yields an asymptotic jet thickness $\delta$. For a wedge with small deadrise angle, one can show that $\delta = \pi V^2 2c/(4 dc/dt)^2 $ \citep{wilson1989mathematics,zhao_faltinsen_1993}. \citet{zhao_faltinsen_1993} took the idea forward and constructed a composite solution (COM) for the pressure that avoids the singularity. Without repeating the derivation, we write their result for pressure distribution on the wedge: \begin{equation}\label{eqn:compositepressure1}
    p = \rho V c \frac{dc}{dt} \frac{1}{\sqrt{c^2 - x^2}} - \rho V c \frac{dc}{dt} \frac{1}{\sqrt{2c (c-x)}} + 2\rho \left( \frac{dc}{dt} \right)^2 \lvert \tau \rvert^{1/2} \left(1+ \lvert\tau \rvert^{1/2} \right)^{-2},
\end{equation}
where $\lvert \tau \rvert$ is a parameter that relates to $x$ via $\delta$ as \begin{equation}\label{eqn:x_c_gap}
    x - c = (\delta/\pi) \left( - \ln \lvert \tau \rvert - 4 \lvert \tau \rvert^{1/2} - \lvert \tau \rvert + 5 \right){,}
\end{equation}
{where the last term from \eqref{eqn:compositepressure1}, and \eqref{eqn:x_c_gap} are identical {to} expressions {already derived by Wagner [more specifically, expressions (10) and (8)  from \citet{wagner1932stoss} respectively].}} As before, the wetting rate $c(t)$ can be obtained from equation \eqref{eqn:wagnerconditionwedge}. The first term in equation \eqref{eqn:compositepressure1} can again be identified from equation \eqref{eqn:wedgetotalpressure} as the slamming pressure, or the outer domain solution. The last term is the inner domain pressure. The second term is the common asymptotic factor in the jet root region, and hence subtracted. The results from this composite solution are plotted as grey solid lines in figure \ref{fig:wedgewagner}. The rise in pressure from the model still exhibits near-singular behaviour, but reaches a finite maximum value. {These values are listed {for different deadrise angles $\beta$} in table \ref{tab:cpmaxcoeffs} alongside the peak pressure coefficients {obtained} from other models {discussed in subsection \ref{sec:OLMMLMGWM}}.} The attainment of {the} impact-peak is followed by the (computed) pressure saturating to a {constant} value {(}of approximately 10$\rho V^2$ {for the deadrise angle studied here)}. The time delay between when pressure on the second sensor starts to rise after the first one is very well reproduced by the model. However the main quantity of interest, the peak pressures, are still over-predicted by the analytical models. Note that the model (equation \eqref{eqn:compositepressure1}) computes the pressure at a single location on the wedge. One {important} reason for the disagreement {is} {therefore} due to the finite size of the transducers' sensing area. Indeed, it was already recognised that instead of the analytical Wagner profile, which is {singular at $x = c$} \citep{takemoto1984some, faltinsen1998water}, a finite size of the sensor in experiments ought to register a space-averaged pressure \citep{zhao1996water,SCOLAN2014470}. Such a quantity {is} better suited to a direct comparison with experimentally obtained peak pressures. {The procedure for the space-averaging is described in Appendix \ref{sec:spaceavappendix}{, and the space-averaged results are presented in figure \ref{fig:wedgewagner}. These will be discussed later together with the models presented in next subsection.}}

{\subsubsection{Original Logvinovich, Modified Logvinovich, and Generalised Wagner models}
\label{sec:OLMMLMGWM}}

\begin{table}
  \begin{center}
\def~{\hphantom{0}}
{
  \begin{tabular}{lcccccccc}
    $\beta \degree$ & OLM & MLM & GWM  &   EIM & ASM & COM & Sensor 1 & Sensor 2 \\[3pt]
       1  & 1925.44 & 3435.80 & 3435.23 & 4044.88 & 4049.18 & 3447.66 & -- & -- \\
       2  & 497.59 & 963.81 & 963.24 & 1009.11 & 1011.68 & 991.21 & -- & -- \\
       3  & 224.18 & 442.70 & 442.13 & 447.25 & 449.18 & 447.10  & -- & --\\
       4  & 126.11 & 249.19 & 248.62 & 250.72 & 252.30 & 251.95  & -- & --\\
       10 & 19.84 & 40.89 & 40.32 & 38.85 & 39.68 & 38.92 & 17.74 $\pm$ 1.54 & 19.79 $\pm$ 2.36    \\
       20 & 4.66 & 10.49 & 9.92 & 8.87 & 9.31 & 8.89 & -- & --\\
       30 & 1.85 & 4.81 & 4.24 & 3.45 & 3.70 & 3.46 & -- & --\\
       40 & 0.87 & 2.78 & 2.21 & 1.63 & 1.75 & 1.63 & -- & --\\
  \end{tabular}}
  \caption{{Values of $C_{p,\text{max}} (\equiv P_{\text{max}}/\rho V^2)$ for {several}   deadrise angles $\beta$ from the {ordinary Logvinovich model (OLM), modified Logvinovich model (MLM) and generalized Wagner model (GWM) as described} in \citet{korobkin_2004} (columns 2--4), numerical solutions to \citet{dobrovol'skaya_1969} by \citet{WANG201723} (EIM, column 5), {the} asymptotic solution {from} \citet{WANG201723} (ASM, column 6), and {the} {composite} {solution} (COM) from \citet{zhao_faltinsen_1993} (column 7).} {{For comparison we have added the m}easured $C_{p,\text{max}}$ from figure \ref{fig:wedgewagner} in columns 8 and 9{, where one needs to realise that these values, other than those of the theoretical models, constitute pressures that are space-averaged over the sensor surface, and therefore considerably smaller than what is found in most models.}}}
  \label{tab:cpmaxcoeffs}
  \end{center}
\end{table}

{Clearly the composite asymptotic model discussed in the previous subsection is only one of many successful attempts to deal with the divergences present in the original \citet{wagner1932stoss} model. Here we also mention the exact self-similar solution in the form of an integral equation from \citet{dobrovol'skaya_1969} and several approximations that {extend} Wagner's formulation by incorporating higher order terms in the Bernoulli equation \citep{korobkin_2004}.}

{As stated before, in} {equation \eqref{eqn:wedgetotalpressure} both the pressure contributions diverge as $x \rightarrow c$. {Also} the ratio of jet to slamming pressure, {namely} $-x^2 \tan \beta / (\pi c \sqrt{c^2 - x^2})$, {diverges to} {negative infinity} {in this limit}. {Generally, the result of the approach taken in \cite{korobkin_2004} is that} the region of interest {is limited} {up to some point $a$,} a short distance before $x=c$. {Following} \citet{korobkin_2004} we write} \begin{equation}
    {a(t) = \xi c(t),\;\; \xi = \sqrt{1-X^2},}
\end{equation} 
{{where} $X$ has the following {functional} forms for the original Logvinovich model (OLM), modified Logvinovich model (MLM) and generalized Wagner model (GWM)
\begin{align}
    X &= \frac{2 \tan \beta}{\pi} \qquad \text{(OLM)} \label{eqn:X_OLM},  \\ 
    X &= \frac{\sin(2 \beta)}{\pi \left[ 1+ \sqrt{1-4 \pi^{-2} \sin^4 \beta} \right]} \qquad \text{(MLM)}, \text{ and} \\ 
    X &= \frac{\sin (2 \beta)}{\pi \left[1+\sqrt{1-4\pi^{-2} \sin ^2 \beta \left( \sin ^2 \beta + \pi -2 \right)} \right]} \qquad \text{(GWM)}
\end{align}
respectively. The corresponding pressures {are equal to} 
\begin{align}
    P(x,t ) &= \frac{\rho V^2}{2} \left[ \frac{\pi}{\tan \beta} \frac{c}{\sqrt{c^2-x^2}} - \frac{c^2}{c^2-x^2} \right] \qquad \text{(OLM),} \label{eqn:OLM} \\
    P(x,t ) &= \frac{\rho V^2}{2} \left[ \frac{\pi}{\tan \beta} \frac{c}{\sqrt{c^2-x^2}} - \cos ^2 \beta \frac{c^2}{c^2 - x^2} - \sin ^2 \beta \right] \qquad \text{(MLM),} \text{ and} \label{eqn:MLM}\\
    P(x,t) &= \frac{\rho V^2}{2} \left[ \frac{\pi}{\tan \beta} \frac{c}{\sqrt{c^2-x^2}} - \cos ^2 \beta \frac{c^2}{c^2 - x^2}  - \sin ^2 \beta + 2 -\pi  \right] \qquad \text{(GWM),} \label{eqn:GWM}
\end{align}
respectively, {where the expressions are understood to hold for $x \leq a(t)\, [< c(t)\,]$.} The maximum (point-)pressure coefficients from these models, alongside the results from {the self-similar integral equation} by \citet{dobrovol'skaya_1969} (using the computed numerical results from \citet{WANG201723}) are compared in table \ref{tab:cpmaxcoeffs}. {Finally, the space-averaged pressure time series resulting from the three models presented above are added to figure \ref{fig:wedgewagner}.}} {From that figure we conclude that, generally, all sensor-averaged models compare well to the experiments; the composite model reproduces the peak values slightly better, whereas the OLM and GWM models are closest to the experiment in the period after the peak, up to the moment that the inner region reaches the knuckle. Similarities and dissimilarities with observations in the literature are discussed in further detail in the next subsection.}

{\subsubsection{Variation of peak pressure along the body}}

Experiments by \citet{chuang1966deadrise}, \citet{chuang1971cones} and \citet{yamamoto1984water} showed for a variety of deadrise angles and velocities that peak pressures varied widely over the surface of impacting cones and wedges. It can be concluded from their vast catalogue of data that the peak pressures followed no systematic trend in how they varied from locations close to the keel to those far from it. Instead, the variation in peak pressures between different locations appeared to result from experimental irregularities such as a non-constant velocity \emph{during} impact or marginally asymmetric impacts.

\citet{PESEUX2005277} performed cone impacts on water with a deadrise angle of 10$\degree$. An attempt to control constant velocity was made. As with our experiments, it was seen that the peak pressures on the sensor farther away from the keel were higher than those on the first sensor. A similar observation was made by \citet{debacker2009experimental} using a cone with a deadrise angle of 20$\degree$. However, using a cone with $\beta =$ 45$\degree$, \citet{debacker2009experimental} observed that the peak pressure on the sensor closest to the keel was higher. \citet{lewis2010impact}, \citet{yettou2006experimental} and \citet{tenzer2015experimental} all used \emph{free-falling} wedges with small deadrise angles, and observed the peak pressures to be higher close to the keel. A freely-falling wedge decelerates upon impact, {which} may thereby result in smaller peak pressures with its increasing immersion.

In contrast, when the wedge is impacted with a controlled `constant' velocity, the moving assembly experiences a large retarding force upon impact. It is conceivable that in a bid to maintain a constant velocity, the assembly accelerates to compensate the loss of momentum. However, since (as shown in appendix \ref{sec:wedgesappendix}) we are able to maintain a constant speed throughout impact, post-impact deceleration in the case of free-falling wedges is not the cause {of} different peak pressures (which was seen throughout earlier figures \ref{fig:conepressures}, \ref{fig:wedgepressures} and \ref{fig:wedgewagner}, where the peak pressure on the second sensor from the keel was consistently found to be higher than from the sensor closer to the keel.) 

Taking into account that the actual quantity measured by the pressure sensor is the force on the sensor’s diaphragm, which is then calibrated towards a pressure using the dimensions of the diaphragm, we directly integrate the theoretically expected pressure distributions from {the models described above (equations \eqref{eqn:OLM}--\eqref{eqn:GWM} and} \eqref{eqn:compositepressure1}) over the sensing area of the transducers to obtain this force. Subsequently we divide through the sensor area to obtain the space-averaged impact pressure over the sensing area of the transducers, which may thus be expected to be equal to the measured pressure. This is done for both the locations along the wedge, where the sensors are installed in the experiment. The results are plotted alongside the non-dimensionalised experimental results in Figure \ref{fig:wedgewagner}. {With the exception of OLM,} the space-averaged pressures can be seen to {closely} follow the rise-characteristics of the measurements. Moreover, the observation that a higher impact pressure is attained on the location farther from the keel (panel (b) in Figure \ref{fig:wedgewagner}) is shown to occur as a result of them being measured over a finite area. {The peak pressure coefficients from space averaged pressures behave in a markedly different manner with $x$ as compared to $C_{p,\text{max}}$ of point-pressures (table \ref{tab:cpmaxcoeffs}), which are constant at all $x>0$ along the body.} {{It is} important to {realise} 
that it is in {fact the} sensors with the same size, that {are} installed at different locations, which 
breaks the self-similarity of the pressure signals that would 
{be expected} from the self-similar theory.} {If one could use sensors with a dimension that increases linearly with the distance to the keel, the sensor signals would be expected to become evidently self-similar again.}

{\subsection{Cone impact pressures}}

In this section we turn to experiments slamming a cone on water. As with the pressure measurements from the wedge (figure \ref{fig:wedgewagner}), the experimental data from cone impacts were obtained over a large range of velocities, and are very well collapsed by inertial pressure and time scales.

\begin{figure}
    \centering
    \includegraphics[width=.99\linewidth]{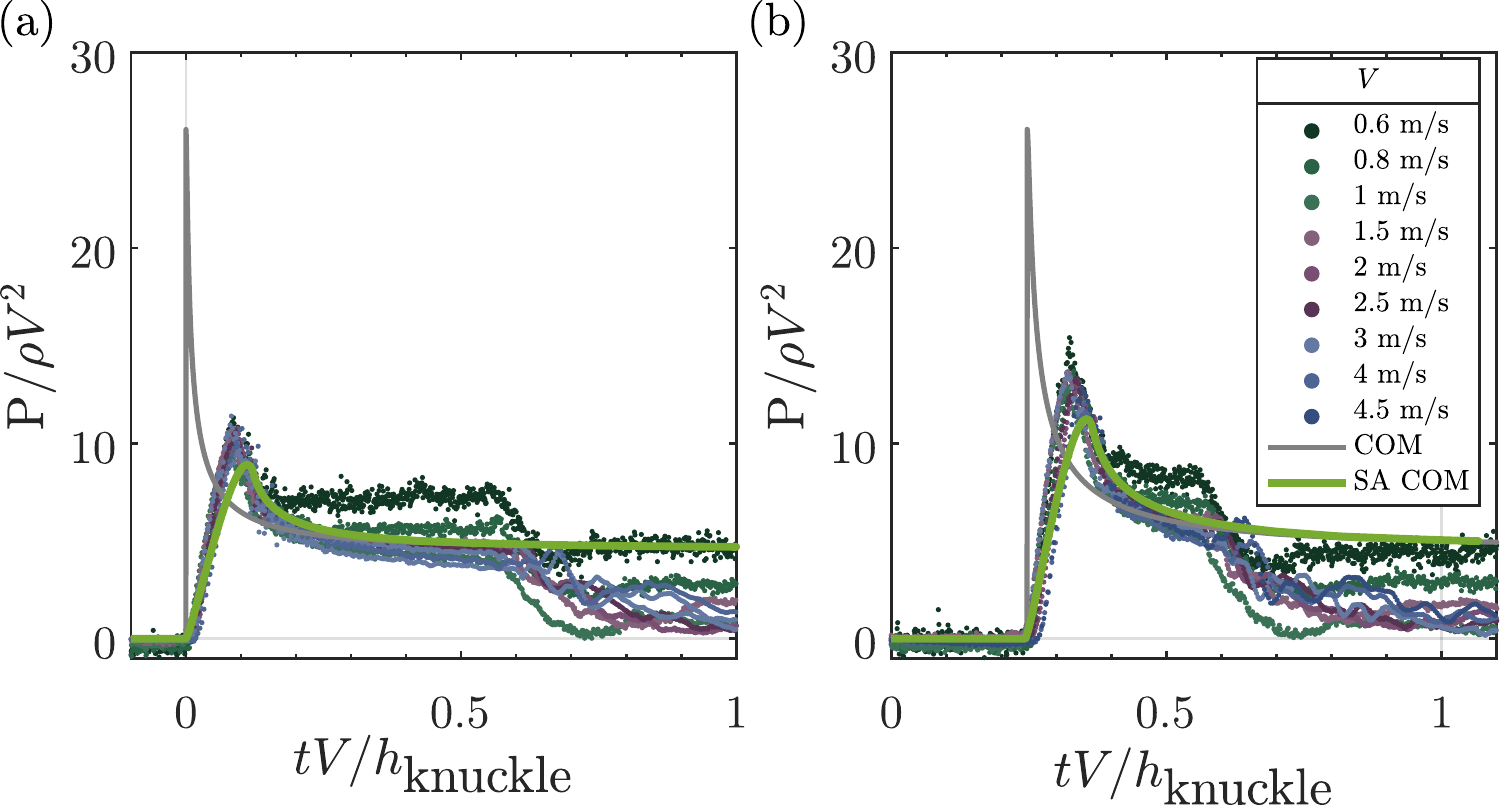}
    \caption{Pressure measurements from the impacting cone at different impacting velocities $V$ are non-dimensionalised and plotted as a function of dimensionless time $t V / h_{\text{knuckle}}$. Results from the first sensor are found in panel (a), while those from the second sensor in panel (b). The quantity $h_{\text{knuckle}} = D \tan \beta /2 \approx$ 6.17 mm. As earlier plots, the time series were shifted so that the point at which the reading on first sensor starts to rise lies at $t = $ 0 ms. The grey solid curves are computations from the composite solution (COM) for the cone (equation \eqref{eqn:compositepressure2}), while the green solid curves are same composite solution, space-averaged over the area of the pressure transducers (SA COM).}
    \label{fig:conemodelcomparison}
\end{figure}

{For a cone impact on water, the flow potential in the domain is known using an axisymmetric solution from \citet{Schmieden1953} and \citet{faltinsen1998water}}: in place of the expanding plate, the no-penetration body condition was transferred to an expanding disc. Thus the treatment remains analogous to the one for a wedge. The remaining boundary conditions were the same as before: $\phi = 0$ on the free surface, and $\partial \phi / \partial y = -V$ for $x<c$, where $x$ now designates the radial position on the cone. The resulting flow potential on the body along the $x-$axis becomes \begin{equation}\label{eqn:coneaxisympotential}
    \phi(x,0) = -\frac{2V}{\pi} \sqrt{c^2 - x^2}, \text{ for } x<c.
\end{equation}
The Wagner condition for cone was calculated to be \begin{equation}\label{eqn:conewagnercondition}
    c(t) = \frac{4 V t}{\pi \tan \beta}.
\end{equation}
\citet{shiffman1951force} found the same result in the outer domain using an elliptical contact line. Next, the axisymmetric solution in outer domain was asymptotically matched with the two-dimensional solution in {the} jet domain to regularise over the singularity at $x=c$. The matching yielded a jet thickness $\delta = V^2 c /2\pi (dc/dt)^2$. Note from equation \eqref{eqn:coneaxisympotential} that with this treatment, only the outer domain solution is modified by a multiplicative factor of $2/\pi$. Since in the jet domain, the two-dimensional solution was used, its pressure contribution remains the same as in equation  \eqref{eqn:compositepressure1}. The composite pressure solution for the cone thus becomes  \begin{equation}\label{eqn:compositepressure2}
    p = \frac{2}{\pi}\rho V c \frac{dc}{dt} \frac{1}{\sqrt{c^2 - x^2}} - \frac{\rho V }{\pi} \frac{dc}{dt} \left( \frac{2c}{c-x}\right)^{1/2} + 2 \rho \left(\frac{dc}{dt}\right)^2 \frac{ \lvert \tau \rvert ^{1/2}}{ \left( 1 + \lvert \tau\rvert ^{1/2}\right)^{2}},
\end{equation}
where $\tau$ is related to $(x-c)$ and $\delta$ as in equation \eqref{eqn:x_c_gap} \citep{faltinsen1998water}. The results from equation \eqref{eqn:compositepressure2} are compared with non-dimensionalised measurements in figure \ref{fig:conemodelcomparison}. 

The {origin of the} time series in figure \ref{fig:conemodelcomparison} were shifted {such that} $t = 0$ {corresponds to the moment at which} the reading on the first sensor from the keel starts to rise. Accordingly, the time delay between when the peak pressures are attained on the two sensors is again very well reproduced by the modified Wagner condition for a cone.

After reaching the peak pressure, the measures at the given locations reduce and approach a saturated value shown by the plateau. At later times ($t  V / h_{\text{knuckle}} \approx $ 0.66), the sudden drop in experimental measurements at both locations corresponds to the liquid jet reaching the cone's knuckles and detaching completely. 

The space-averaged pressures from the composite solution for a cone (equation \eqref{eqn:compositepressure2}) are computed for the two locations where the sensors were installed in the experiment. They are compared with the non-dimensionalised measurements in figure \ref{fig:conemodelcomparison}. As with the wedge, the agreement is found to be remarkably good at both the sensors. However, compared to the results for a wedge from figure \ref{fig:wedgewagner}, on the cone, the peak pressure is slightly under-predicted, while the rise time is somewhat over-predicted.\\

%
%

\subsection{Comparing peak pressures on cones and wedges}
\label{sec:3dfloweffect}

The most notable progress in modelling water-entry pressures has been made using two-dimensional models \citep{wagner1932stoss, logvinovich1969hydrodynamics, howison_ockendon_wilson_1991, zhao_faltinsen_1993} and their extensions {in three dimensions or cylindrical coordinates} \citep{scolan_korobkin_2001,faltinsen2005generalized,oliver2002water, moore2014new, moore_howison_ockendon_oliver_2012, 10.1093/imamat/hxu026}. However in practice, all impact processes are three-dimensional. As such, it is important to know the limits of how the Wagner treatment is extended to model the wetting rate in a three-dimensional system. The most straightforward extension to three-dimensions is made by considering an axisymmetric impact. In this context, the impact of a cone represents an important test case of how the Wagner condition on a wedge is modified to include three-dimensional flow effects. It can be seen from equations \eqref{eqn:compositepressure1} and \eqref{eqn:compositepressure2} that the rate of local rise-up of water along the impactor body has a crucial contribution to the impact pressure. 

\begin{figure}
    \centering
    \includegraphics[width=.85\linewidth]{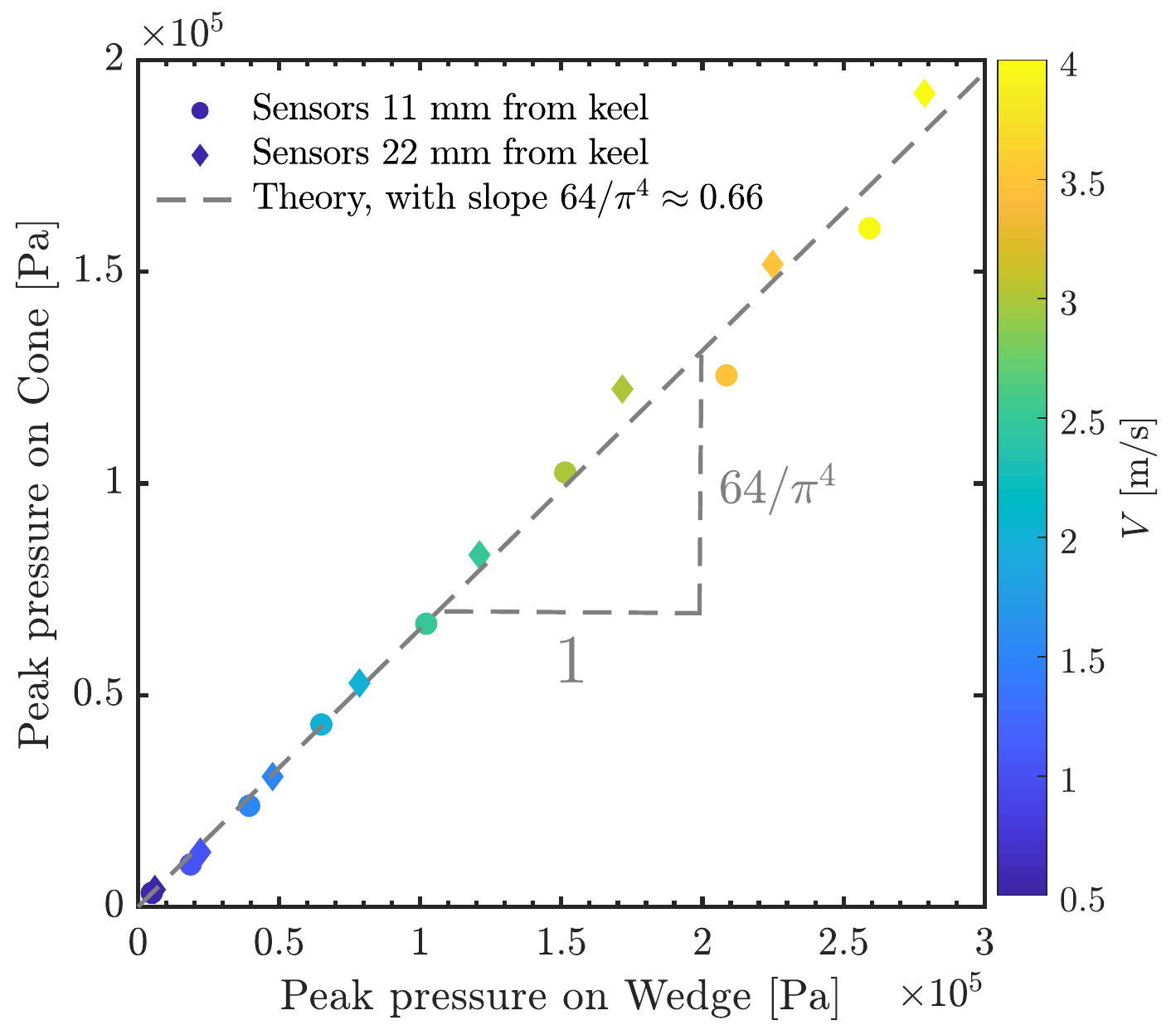}
    \caption{The peak pressures from wedge and cone are directly compared at both sensors. The peak pressures measured on the cone are found to be {$64/\pi^4$} times to those measured on the wedge across the whole range of $V$ used. The measurements are compared to the analytical result (equation \eqref{eqn:peakpressureratio}) finding excellent agreement.}
    \label{fig:wedgecone_peaks}
\end{figure}

The maximum pressure from {the} model {at any given location along the impactor} is simply $\rho {(dc/dt)}^2/2$. This is clearly {larger than what any finite-sized sensors would measure} (figures \ref{fig:wedgewagner} and \ref{fig:conemodelcomparison}). However, the relative magnitudes of peak pressures between a wedge and cone from $\rho {(dc/dt)}^2/2$ may be compared to those measured in experiments. A direct comparison of peak pressures between cone and wedge from sensors at the same horizontal locations is made in figure \ref{fig:wedgecone_peaks}.

Recall that for a wedge, $c(t) = \pi V t /2 \tan \beta$ (equation \eqref{eqn:wagnerconditionwedge}). While for a cone, $c(t) = 4Vt/\pi \tan \beta$. The ratio of peak pressures is therefore expected to be \begin{equation}\label{eqn:peakpressureratio}
   \frac{ p_{\text{peak}}^{\text{cone}} } { p_{\text{peak}}^{\text{wedge}}} = \frac{64}{\pi^4} {\approx 0.66}.
\end{equation}
This ratio is plotted as the dashed line in figure \ref{fig:wedgecone_peaks}, and it agrees very well with the experimental data. This {is an independent measure, corroborating the inclusion of} three-dimensional effects only in the outer domain, by modifying the wetting rate $c(t)$. Note that, { \citet{chuang1969theoretical} had found the same ratio {of peak pressures} to be $0.75$. {Although peak pressure data for wedge and cone impacts are} certainly present in \cite{takemoto1984some}, {unlike in the current work,} it was not analysed beyond the documentation of the pressure peaks.} \\

{{Finally, we turn to the decay of the pressure after the occurrence of the peak} 
in figures \ref{fig:wedgewagner} and \ref{fig:conemodelcomparison}{, that is, after the root of the jet has passed the pressure sensor.} 
While for wedges the experimental pressure consistently decays faster than that of the models, this is not the case for our pressure measurements on the cones{, where experiments are coinciding or even slightly above the theoretical prediction. Whereas the origin of this observation is unknown, it does appear that the deviations from the models in both cases become smaller with larger impact velocities. We therefore speculate that they could well find their origin in (viscous) interaction of the liquid with the impactor body, in both the outer and the inner region. This is also consistent with the fact that for the wedge impacts the kink (at $t V /h_{\text{knuckle}} \approx 0.5$ for wedges and $\approx 0.6$ for cones) that corresponds to the jet root passing the knuckle becomes more and more pronounced for increasing impact velocities.}}


\section{Air cushioning before {cone} impact}

\begin{figure}
    \centering
    \includegraphics[width=.65\linewidth]{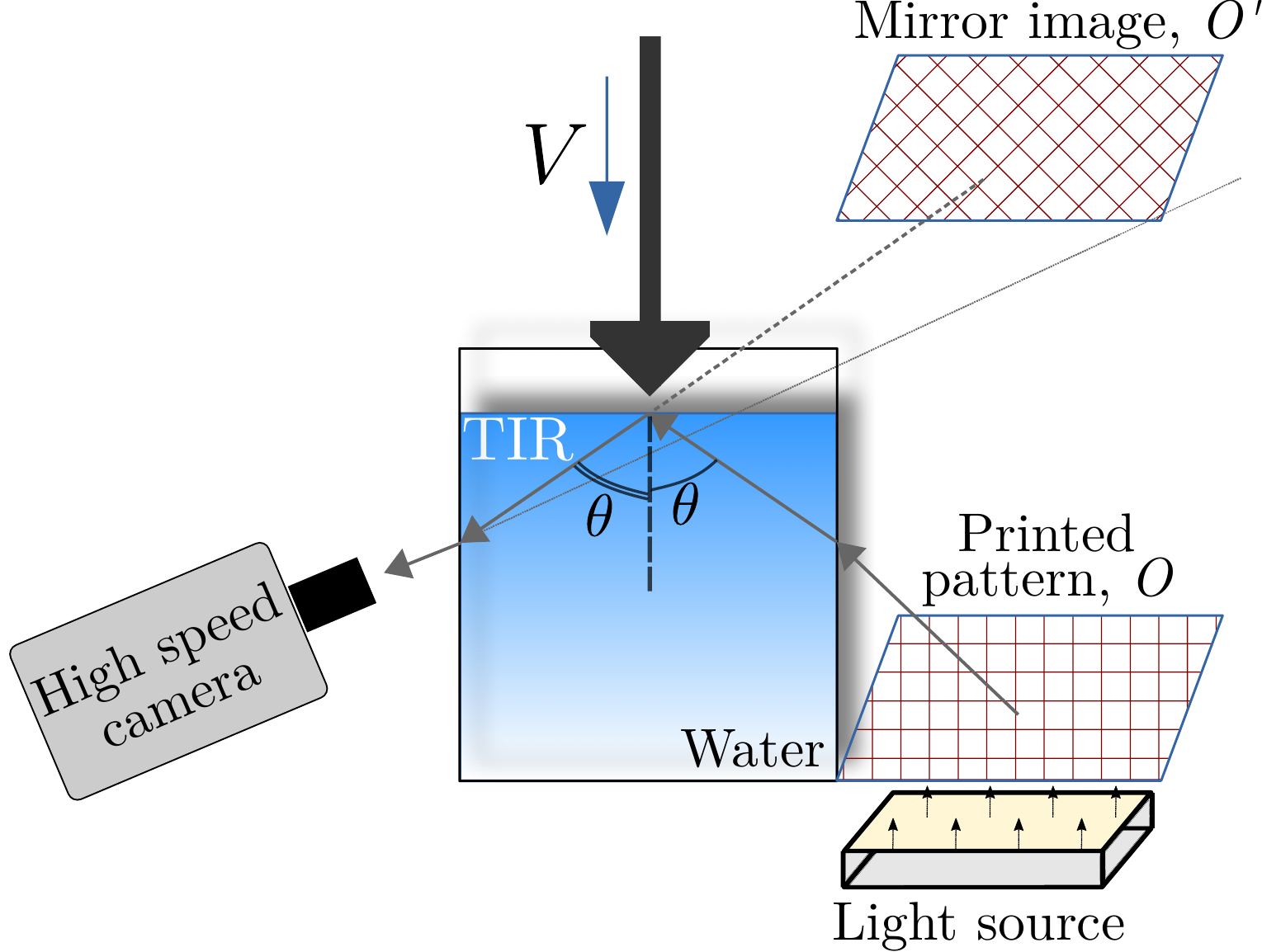}
    \caption{Total-internal-reflection deflectometry setup \citep{tirdmanuscript} {reflects the state of the water surface when it deforms in response to an external disturbance. The water surface's movements are interpreted as a deforming mirror. When an object touches the water interface, the reflecting surface disappears in those regions and turns dark.}}
    \label{fig:cone_tirdsetup}
\end{figure}

\begin{figure}
    \centering
    \includegraphics[width=.99\textwidth]{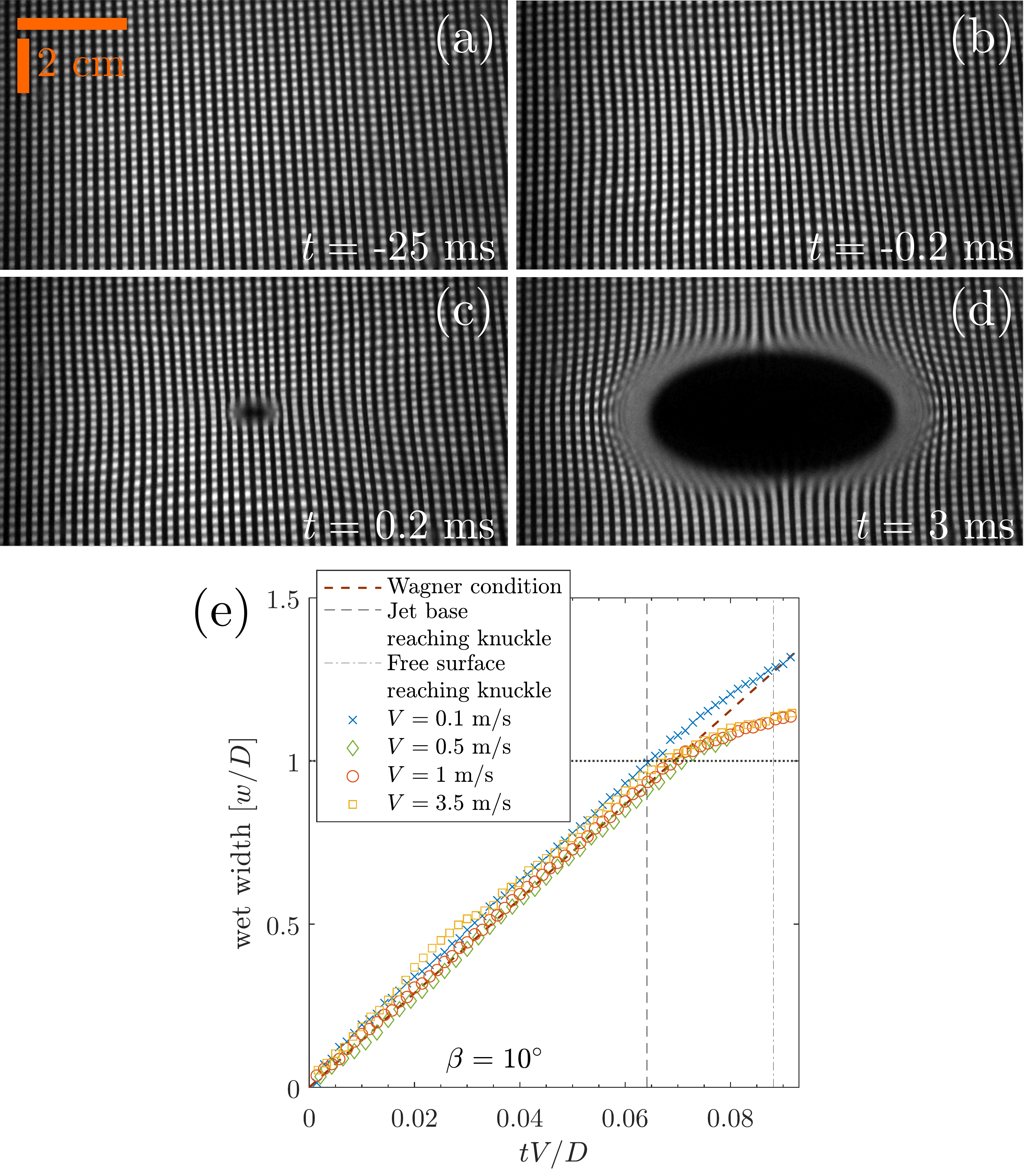}
    \caption{Snapshots of the water surface through different stages of a 70 mm wide cone with $10\degree$ deadrise-angle entering water at $V=1 $ m/s. The cone makes the first contact with the water surface at $t=$ 0 ms. Notice in panel (b) at $t=$ -0.2 ms how the pattern is deformed due to air flow, \emph{prior} to the cone coming into contact with the water surface. Note that the length scales differ along {horizontal and vertical} directions due to the reflected image at water surface being {expanded} along the {horizontal} direction. A video of the process shown is available in supplementary material. (e) {{Dimensionless w}etted width {$w/D$ as a function of dimensionless time $tV/D$} {measured from the total width of the dark section from post-impact images such as in panels (c--d)} from TIR view, and compared with Wagner condition \eqref{eqn:conewagnercondition}.} {The dotted horizontal line is drawn to show the width of the cone. The dashed vertical line shows the time when {the Wagner condition \eqref{eqn:conewagnercondition} predicts the jet base to reach the knuckle of the cone} and detach. The vertical dash-dotted line is the time taken for the cone to traverse its actual depth ($h_{\text{knuckle}}$) until the stationary water level{, i.e., $tV/D = h_{\text{knuckle}}/D$.}}}
    \label{fig:10degtirdraw}
\end{figure}

{Any object that is close to impacting on water makes it presence felt to the target liquid via an intervening air cushioning layer. Here,} as the cone approaches the moment of impact, it pushes the ambient air from between itself and the target liquid surface. The changes in pressure in this air layer has the effect of deflecting the water surface away from the approaching impactor under its center. {This stage is of importance to impact loading as it can entirely change the stages of first contact between the impactor and the free surface. One obvious way in which it effects the impact pressures is at the keel - where the relative impact velocity is reduced due to the free surface moving in the same direction as the approaching impactor. This is the reason why the Rankine water-hammer pressure $\rho c_a V$ (with $c_a$ the acoustic speed in water), will {highly unlikely} be measured at the keel. Another equally important reason is that air cushioning can cause an air film to be entrapped under the cone.} {\citet{chuang1970investigation} reported from visual inspection that air cushioning effect is} most prominent when the deadrise angle is less than approximately 3$\degree$. {An air film spanning the cone's entire width is entrapped when the radial slope of the water surface from the cone's symmetry axis to its edge is larger than the cone's own slope. Alternatively, a film of lesser width will be entrapped when the same criterion is satisfied over a smaller radial distance. Clearly when {such a} film will be entrapped under a finite-deadrise-angle body, Wagner {pressure theory} will not be applicable to the regions covered by the air film. In such a situation, even an area-integrated Wagner theory will also not be {able} to {reliably} estimate the impact forces.} {And, even in the case that no air film is entrapped, the velocity and acceleration acquired by the deforming liquid surface prior to impact may influence the magnitude of the measured pressure peaks, especially very close to the keel.}   

{Whether the air film is entrapped or not, its thickness is determined by the extent of vertical deflection that the water surface undergoes due to the air cushioning layer. This depth of the free-surface cavity is created due to the stagnation point set up by the air layer being squeezed out. The stagnation pressure's magnitude is determined by the inertial scales in the problem. We have shown in \citet{jainkhPRF} using a flat disc (or zero deadrise angle cone), that the deflection magnitude can be well estimated from potential flow considerations. However it is not clear how much varying the deadrise angle will affect the stagnation pressure. Since the expected movement of the liquid surface are {too} small to be seen from a side view, and blocked from the top view by the impactor itself, we record the water surface's movements using a total internal reflection setup.}

{We perform the experiments in the same transparent tank as earlier described in section \ref{sec:conewedgesetup}. The free surface is visualised by placing a camera such that it makes a large enough angle ($\arcsin n_a/n_w$, where $n_a$ and $n_w$ are the refractive indices of air and water respectively) with the normal vector of the water surface, so as to satisfy the requirements of total internal reflection. The water surface's movements are observed in such a setup by means of a reference pattern such as shown in figure \ref{fig:cone_tirdsetup}. Thus the water surface acts as a deforming mirror in this configuration.}

{Using {such a total-internal-reflection (TIR) setup,} we can visually inspect the free surface throughout a cone impact. One such example is {shown in figure \ref{fig:10degtirdraw} for }a $10\degree$ deadrise angle cone (also see movie 1). {Additional examples showing air cushioning under a 1 and 2 $\degree$ deadrise angle cones are discussed in supplementary material and {movies 2 and 3}. {We show in {movie} 2 that (and also supplementary material) that} air becomes entrapped under the {1}$\degree$ cone, with its keel only starting to wet approximately $1.13$ ms 
after the cone first comes into contact with the water along its edges. From {movie 3} we can see the entrapment already ceases to occur with $\beta=2\degree$ cone.} It is clear that air cushioning stays relevant to the extent that an air layer can be fully entrapped under a cone. The entrapped air film's final fate, whether how it may be fragmented or expelled from under the cone, is determined by stagnation pressure in the liquid after impact. The mechanisms for post-impact evolution of the air film would be the same as described in case of a flat disc in \citet{jain2020air}.}

{{The same TIR setup used for visual inspection is subsequently used to quantitatively} resolve the {deformation of the water surface for cones with a deadrise angle} up to $30 \degree$.} {For these measurements, we fabricated additional cones of $D=70$ mm, with deadrise angles 5--30$\degree$, each printed using Formlabs Clear V4 resin.} {We show in the supplementary material that using these different materials (PET or 3D printing) to produce the cones does not significantly modify their surface finish, such that the air flow that causes cushioning underneath them is significantly affected by the changed surface roughness.}

{Since the thickness of an entrapped air film is determined by the pre-impact cavity along the free surface,} {here we seek to explain the growth and the magnitude of this transient cavity that air-cushioning creates under bodies with non-zero deadrise angle. It is known that the initial thickness of this air layer under a disc (or a 0$\degree$ cone) at moment of impact is mainly determined by the stagnation pressure on the water surface along its symmetry axis $(r=0)$ \citep{jainkhPRF}. The stagnation pressure in turn is fully determined by inertial scales in the problem \citep[Chapter 3]{ujthesis}, impactor size $D$ and velocity $V$.}  

\subsection{Measuring the free surface deflections using TIR-Deflectometry}

The free surface of the water in the tank is {observed} using a {total {internal} reflection (TIR)} configuration as shown in figure \ref{fig:cone_tirdsetup}. A fixed pattern ($O$) placed on one side of the tank is reflected at the water surface, and recorded by a camera on the other side. The image appears to come from `behind' the reflecting surface ($O'$). When the water surface moves from its stationary position, the mirror in the present optical setup moves, and distorts the pattern seen by the camera. When an object is present at the water surface, it obstructs the mirror such that the water surface at all such locations does not reflect the pattern.

Some examples of the free surface being visualised in such a setup in an experiment with a cone impacting on water are shown {in figure }\ref{fig:10degtirdraw} {{and movies 1--3}}, wherein the first contact occurs at $t=0$. In particular it can be seen at {times before impact} that the pattern is distorted, {already} indicating that the water surface has undergone some {deformation} before the cone has come into contact with it.

Such movements of the pattern are recorded using high speed imaging, and quantified using a PIV algorithm. The stacks of displacement fields thus obtained correlate to the local spatial gradients of the water surface. {The displacement fields are integrated in a least-squares sense to obtain the instant-wise reconstruction of the {deformation of the} water surface.} The details {{of this} reconstruction are described} in \citet{tirdmanuscript}. {The fully reconstructed free surface is also shown in movie 4.}

{{In addition to the surface reconstruction before impact,} the extent {to which the cone is wetted after impact} is also easily measured {using the TIR view.} Measurements of the wetted region of a cone with $\beta =10 \degree$ {at different impact velocities} are shown in figure \ref{fig:10degtirdraw}(e). They compare very favourably with Wagner condition for cone \eqref{eqn:conewagnercondition}. {Results from similar} measurements with cones with $\beta=$ 5, 20 and 30$\degree$ are available in the {supplementary material}.} 

\subsection{Two-fluid boundary integral simulations}
{We use an in-house boundary-integral {(BI)} code to simulate the interface's response to the pressure build up in the air layer between the cone and liquid. The implementation has been described in detail in earlier works \citep{bergmann_2009,oguz_prosperetti_1993,gekle2011}, and here will only be briefly discussed. We first describe the basic principles.}

{The implementation is based on the flow being inviscid and incompressible. Thus, a scalar flow potential describes flow in the bulk, and the fluid(s) satisfy Laplace's equation. As is done in boundary element methods, the Laplace's equation is transformed using Green's third identity to obtain a so-called boundary-integral equation. This equation consists of only surface-integrals wherein flow potentials are defined only along the boundaries of the concerned fluid domain. From this integral, flow information in the fluid bulk can be deduced. The boundaries of each domain are populated by `nodes' at which the flow potentials are defined. We use an axisymmetric formulation of such a code, which further reduces the surface integrals in the boundary-integral equation to line integrals. The surface then is initiated at the first constituent node of this curve, which in our setup lies on the symmetry axis (at $r=0$). In far field, all surface potentials decay to zero in the formulation. Since we are only concerned with the region nearer to $r=0$, and contribution of far regions to the flow potentials would be negligible, each fluid surface in the code can be cut off at some sufficiently large $r$. Finally, the BI implementation can be made a fully two-phase coupled formulation by incorporating a Laplace pressure jump across the interface, and imposing a balance of normal velocities at the fluids' interface \citep{gekle2011,gordillo07,rodriguez_gordillo2006}.}

{The present simulations are performed with the fully-coupled, two-phase BI code{, that has previously been} used by \citet{peters2013} and \citet{jainkhPRF}. Here the two fluid phases are water and air - both of which are characterised by their respective {densities} and interfacial tension. The cone is defined as a region bounded by solid curves along which the normal flow potential in surrounding air (in respective frame of reference) is set to zero.}

{The node distribution along the interface is initialised such that there are two regions of high node density: the node density gets linearly denser closer to $r=0$, and in the vicinity of $r=D/2$. At the initialisation, the minimum inter-node distance used in the simulations here (and in supplementary material) are typically 0.0005 and 0.001 of the representative length scale (for these simulations, $D/2$). Further the node density evolves in time based on interface curvature and gas velocity, such that the it is allowed to grow past the maximum resolution defined at initialisation. However we impose that the number of nodes between each time step do not grow faster than a multiplicative rate of 1.1. Due to a lack of natural damping in BI simulations, numerical instabilities may accumulate. This is handled by regridding the entire curve every few time steps, and making the new nodes lie exactly halfway in between previous ones. However, some accumulation of numerical instabilities at $r=0$ at later times is not avoided despite the regridding. More details may be found in \citet{gekle2011}.}

\begin{figure}
    \centering \includegraphics[width=.7\linewidth]{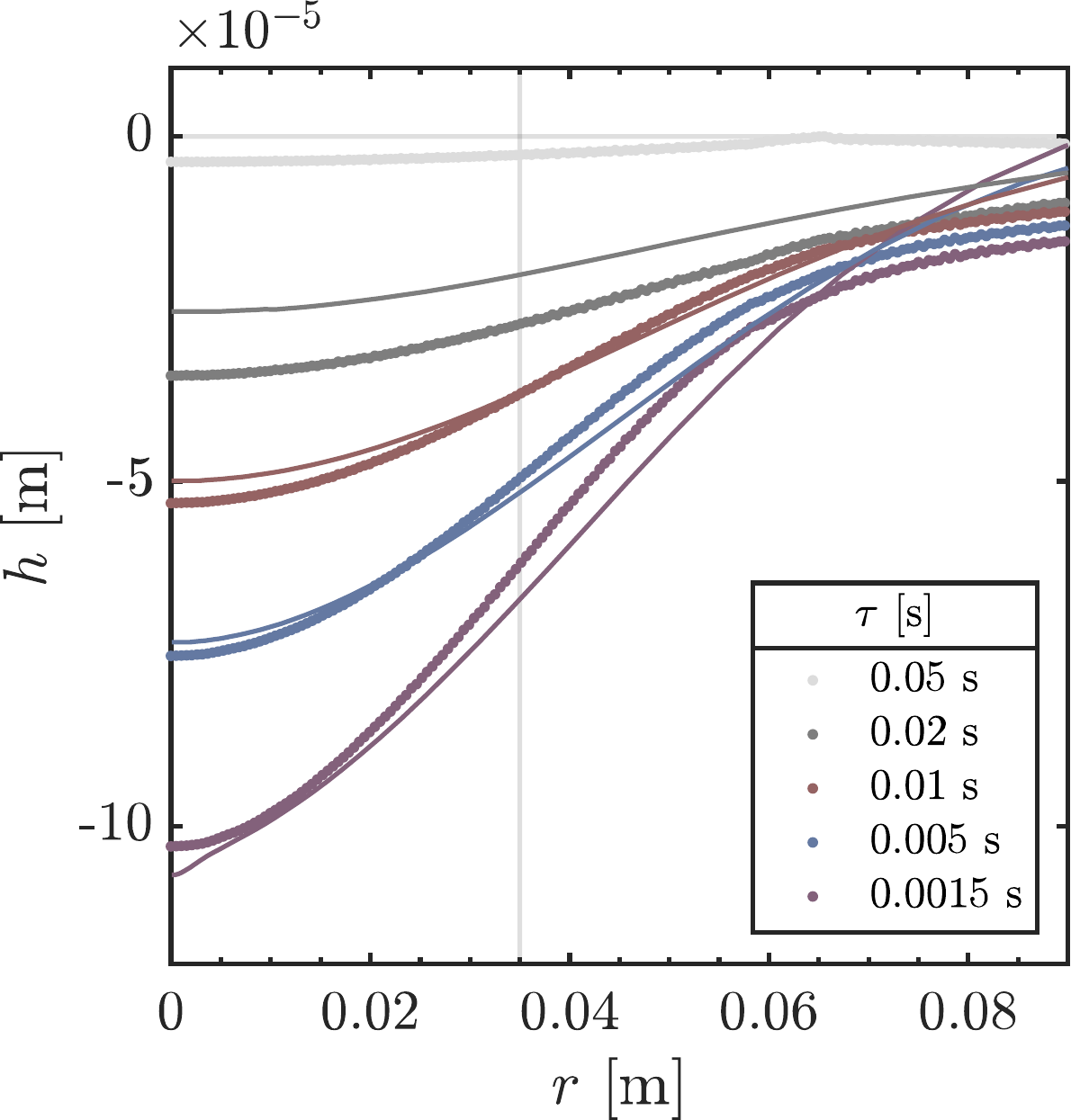}
    \caption{{Vertical d}eformation $h(r,t)$ of the water surface due to air cushioning before the impact of {an approaching $10\degree$ deadrise-angle} cone, as a function of the distance $r$ to the symmetry axis, at several instants in time. {Here, $\tau = t_{\text{impact}} - t$ denotes} the amount of time {remaining until} impact (which {thus} occurs at $\tau=0$). The measured water surface profiles are azimuthally averaged. At $r=0$ (directly under the keel), there is a stagnation point which pushes the water surface {downwards}. The vertical gray line at $r =$ 35 mm {indicates the radial location of the knuckle, i.e.,} where the cone's contour turns away sharply. {Surface profiles {$h(r,t)$ from the} BI simulation are shown as solid lines, {where the colours correspond to those of the} experimental results. The animations for this experiment are shown in movies 4 and 5.}}
    \label{fig:conecushioning}
\end{figure}

\begin{figure}
    \centering
    \includegraphics[width=.99\linewidth]{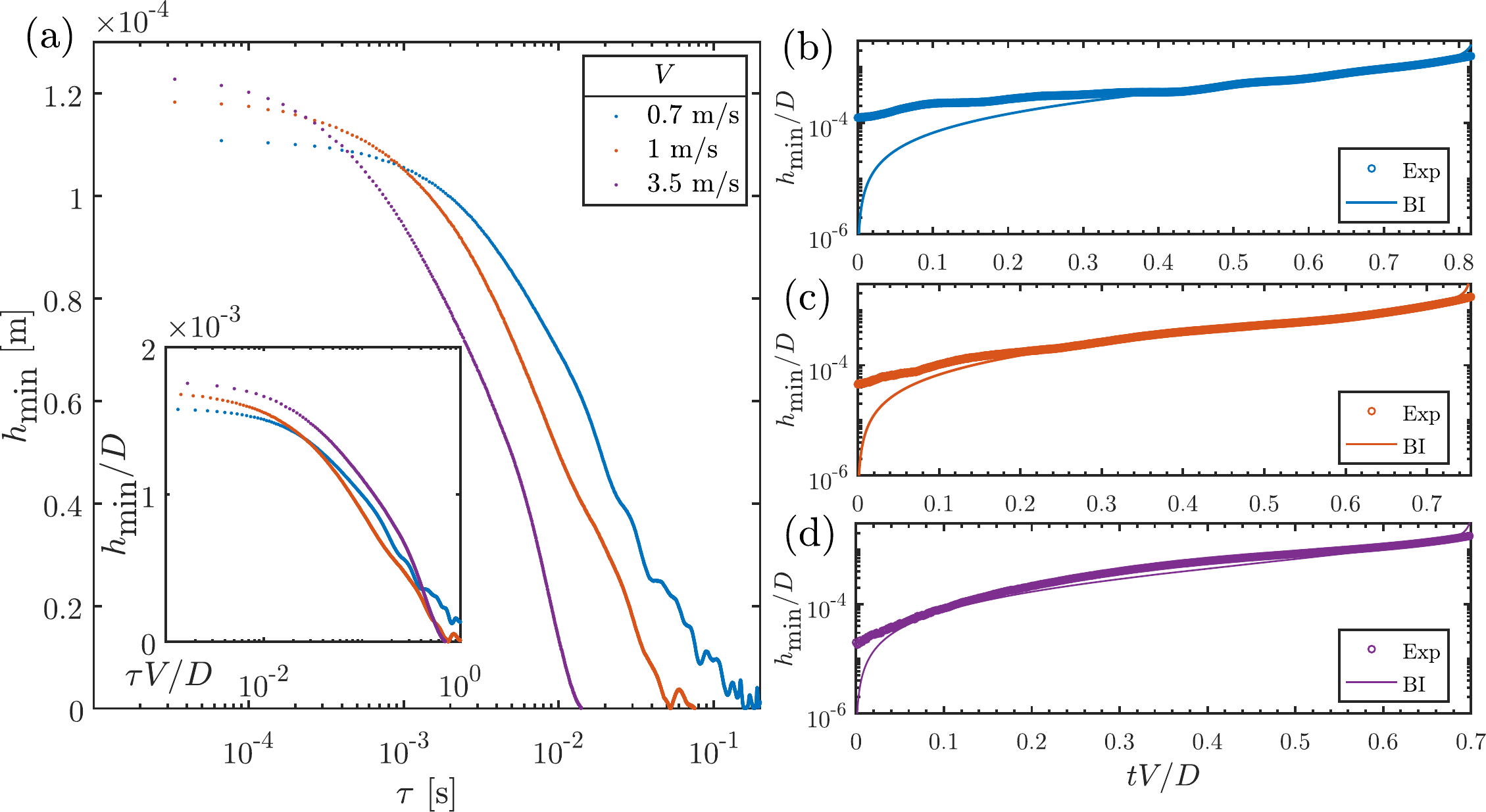}
    \caption{{(a) {Experimentally measured central depth $h_{\text{min}}$ of the deformed} water surface at $r=0$ under the approaching $10\degree$ {deadrise-angle} cone is plotted {versus the time remaining until impact}, $\tau = t_{\text{impact}} - t${, for three different impact velocities $V$.} {The inset shows the collapse of the same data when} nondimensionalised using inertial scales. In panels (b--d) we compare the results from panel {(a) with $h_{\text{min}}$ determined from} two-fluid boundary simulations, {now using forward time $t$. After a short start-up period, the experimental (open symbols) and numerical data (solid lines) data coincide. {Note that in (b--d) the same colour coding is used as in (a).}} }}
    \label{fig:10degcushioning}
\end{figure}

{The simulation is initiated at a non-dimensional time which is determined {as the instance in experiments} when the interface starts to move {downward steadily (see $\tau \approx 0.052$ s in movies 4 and 5 for the $\beta=10\degree$ cone with $V=1$ m/s).}} {At the determined starting time, the interface in the experiment is already in the process of deforming (since even prior to the action of air-cushioning layer, there were unavoidable minuscule fluctuations of the interface), whereas in the simulation, the interface position is initiated at zero. This is the reason for the differences between experiments and BI at early times in figure \ref{fig:10degcushioning}(b--d), which for the present discussion we denote as the start-up discrepancy. The discrepancy becomes smaller as the process progresses in time - this is further discussed in next subsection.} 

\begin{figure}
    \centering
    \includegraphics[width=.6\linewidth]{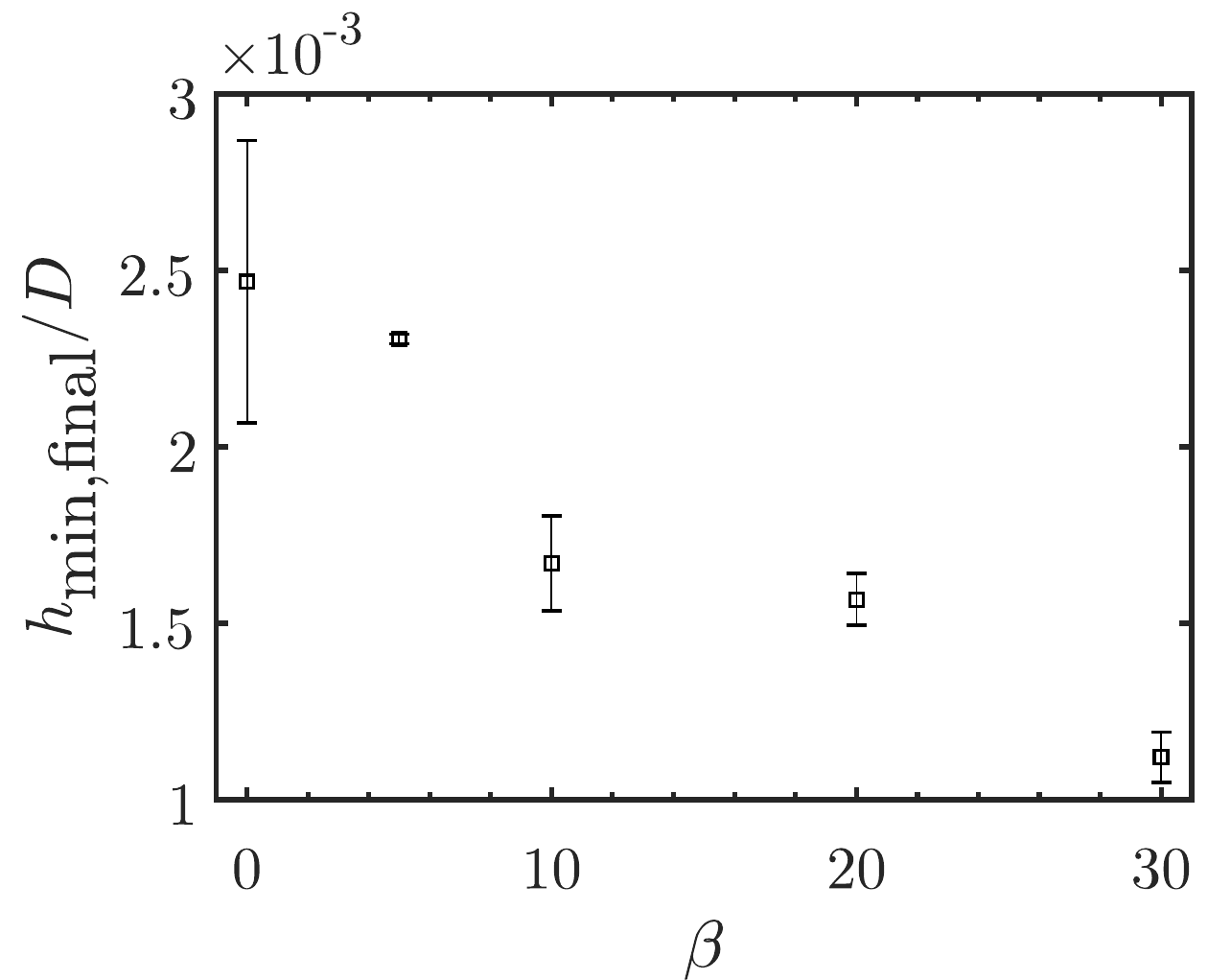}
    \caption{{{The non-dimensionalized final depth $h_\text{min,final}$ of the surface deformation} 
    due to air cushioning is {plotted versus the} deadrise angle for {a disc and several cones} with diameter $D=70$ mm and impact speeds $V$ varying between $0.7$ m/s and $3.5$ m/s. {The error bars denote the variation occurring for different impact speeds.} }}
    \label{fig:hminfinal_beta}
\end{figure}

\subsection{Results}

The azimuthally averaged, reconstructed profiles $h(r,t)$ of the water surface from an experiment {with a cone with a deadrise angle of $\beta=10\degree$ at an impact speed} $V = 1.0$ m/s are shown in figure \ref{fig:conecushioning} {for different moments in time, measured in the amount of time remaining until impact, $\tau = t_\text{impact}-t$.} The water surface {below the keel of the cone (at radial coordinate ${r=0}$)} is pushed down due to {the formation of} a stagnation point at the water-air interface. {Clearly, as the cone approaches the free surface, the deformation becomes larger. Remarkably, the deformation extends beyond the limits of the cone, which is indicated by the vertical gray line. In the same figure we compare the experimental profiles to the numerical BI simulation results at the same instants in time. Whereas initially there is a significant discrepancy between experiment and numerics due to the mentioned start-up effect in the simulation, at later times numerics and experiments agree fairly well.} 

{Subsequently, we plot the time evolution of the depth $h_\text{min}$ of the free surface at the minimum occurring at the center ($r=0$) of the deformed interface in figure \ref{fig:10degcushioning}(a) for three different values of the impact speed $V$. Non-dimensionalizing both time and depth using inertial scales (length scale $D$ and time scale $D/V$) leads to a collapse of the data as shown in the inset of figure \ref{fig:10degcushioning}(a), confirming that the deformation largely follows inertial scaling.} {(In fact, one may argue that the appropriate length scale is obtained by multiplying $D$ with the density ratio $\rho_g/\rho_l$.)}  

{From both figures \ref{fig:conecushioning} and \ref{fig:10degcushioning}, it is clear that the water surface deformation due to gas inertia is appreciable {as early as} when the cone-water separation is of the order of $D$, clearly {indicating that deformation starts long before viscous (lubrication) effects may start to play a role.} {Since during most of the air-cushioning process the {deflection of the interface is very small compared to the gap width between the undisturbed free surface and the impacting cone, i.e.,} $h \ll D$, and {similarly for the} velocity $\dot{h} \ll V$, we may estimate} {the radial gas velocity in the gap as}
\begin{equation}\label{eqn:airvelundercone}
    V_{r,\text{gas}} = \frac{r}{2} \frac{V}{V\tau + r \tan \beta}\,,
\end{equation} where $\tau = t_\text{impact} - t$ denotes the time remaining until impact.}
{The basic mechanism controlling the deflection of the liquid interface under the impactor is described {as follows}: the air in between the impactor and the liquid surface is squeezed out until the moment of contact. As the radial pressure distribution in the air layer is set up, there is a balance between the static and dynamic pressures. At $r=0$, the dynamic pressure is 0, while the static (stagnation) pressure is maximum. Whereas at $r=D/2 $ under the impactor edge, $V_{r,\text{gas}}$ and hence the dynamic pressure are at their maximum. With decreasing $\tau$ until impact, the stagnation pressure at $r=0$ keeps growing and pushes the water surface down. This mechanism has also explained air cushioning under a flat plate \citep{verhagen1967impact,wilson1991mathematical, peters2013,jainkhPRF}. {In figure \ref{fig:10degcushioning}(b--d) we directly compare the experimental curves of figure \ref{fig:10degcushioning}(a) with the corresponding BI simulation result. Again, after a start-up phase, experiments and simulations are in good agreement. It is due to the steady pressurisation of the air layer, and hence the interface, that the discrepancy between the simulations and experiments diminishes at later $t$.} More such comparisons for cones of 5, 20 and 30$\degree$ deadrise angles are discussed in supplementary material.}

It can be seen that the final value to which the {deforming} free surface saturates before the moment of impact is approximately $1.75\cdot 10^{-3}D$. This is somewhat smaller, but of a comparable magnitude to the case of a flat disc, where the same quantity was found to be approximately $2.5\cdot 10^{-3}D$ \citep[Chapters 2--3]{ujthesis}. Any deadrise angle greater than about 3$\degree$ has been found to be sufficient enough to avoid any air being trapped under the impactor \citep{chuang1966deadrise, chuang1971cones, okada2000water}. We also find no air being trapped under the cone despite the substantial deflection of the free surface (see snapshot at $t=$ 0.2 ms in figure \ref{fig:10degtirdraw}). Thus we conclude that air-cushioning effect under a cone with a large deadrise angle, by delaying the moment of touchdown by several milliseconds, and by locally deforming the interface, may serve to slightly soften the impact at the keel. {{From time-traces such as in figure \ref{fig:10degcushioning}, we estimate the change in relative impact velocities to be reduced by 0.1--0.3 \% for our cones with $\beta = 10-30\degree$, and approximately 0.6 \% for the cone with $\beta=5\degree$ (the time series for $\beta$ other than $10\degree$ are shown in supplementary material). Clearly since the impact pressure $\sim V^2$, the reduction in force on account of air cushioning is negligible.} Additionally,} since the deflection is smaller than 2\% of the cone diameter, it is not expected to have any significant effect on the impact pressures further away from the keel. 

{Another measure that can be derived from the data is the angle which the water surface makes with the horizontal at the impact moment. These can be immediately derived from figure \ref{fig:hminfinal_beta}: ranging from $0.286 - 0.127$ {degrees} for $\beta = 0-30 \degree$ deadrise angle cones, each with diameters of 70 mm.} {The impactor will entrap an air layer when this angle that the water surface makes with the horizontal is larger than the impactor's deadrise angle. However, this condition can be favoured if the escaping air reaches a {sufficient} velocity at some $r$ to initiate a Kelvin-Helmholtz instability to cause upwards suction of the liquid, or indeed reach a Mach number of 0.3 to become compressible and thus effectively choke at some finite $r$. For either of the cases, the minimal criterion can be obtained by substituting the critical air velocity as $V_{r,\text{gas}}$ in equation \ref{eqn:airvelundercone}}{; either of these effects ought to favour the upwards suction of the liquid towards the impact at some finite $r$ to enhance the chances of air entrapment.}

{\subsection{Inertial vs. viscous air cushioning} }

{When compared to the relevant length and timescales conveyed by the viscous air-cushioning models \citep{rosshicks2019} for the same impact geometry, our air-cushioning results present {a} very different picture. The model in \citet{rosshicks2019} explicitly starts with the assumption that gas pressure buildup in the intervening layer does not influence the liquid until the vertical separation between the solid body and the liquid free-surface is of thickness $\varepsilon^2 D$, where $\varepsilon \equiv( \mu_g/(\rho_l V D) )^{(1/3)}$, {$\mu_g = 1.81\cdot10^{-5}$ Pa$\cdot$s} is the gas {dynamic} viscosity, {$\rho_l = 998$ kg/m$^3$} the liquid density and $V$ the approach velocity. In our experiments, $D = 0.07 $ m. Using $V=0.1$ m/s, this separation $\varepsilon^2 D$ is 13.2 microns, while for 1 m/s we obtain $\varepsilon^2 D=2.84 \mu$m. 
These body-liquid separations are {much smaller than those of} 
our observations in panels (b--d) of figure \ref{fig:10degcushioning}, where the body-liquid separation is approximately 5 cm when the interaction is clearly established. Similar results are shown in supplementary material for cones with {deadrise angles} $\beta = $ 5, 20 and 30$\degree$. Assuming that viscous-pressurisation of the gas layer indeed begins when $\tau = \varepsilon^2 D/V$, would imply that, of the distance of $0.7-0.8D$ (approximately 5 cm) that the $10\degree$ cone travels at $V = 1$ m/s whilst interacting with the liquid surface via the air layer,  the viscous cushioning is active for only 0.0057\% of the initial gap distance. Thus, the deformations caused by viscous air cushioning will be created on an interface that has already been substantially deformed into a cavity by the earlier inertial air cushioning. Another important aspect in which the two will differ is the actual volume of the created cavity. While potential flow creates an interface cavity of the width of the impactor, the lubrication-pressure distribution would result in interface deformations that are much more localised around $r=0$. The natural limits to when potential or viscous air cushioning would hold for a spherical projectile, were addressed using both numerics and theory by \citet{bouwhuisinitial}.}

{In appendix \ref{sec:inertialviscouscushioning}, we use the results from \citet{bouwhuisinitial} to 
{approximate the crossover} air-layer thickness {where} the type of air cushioning, in our experiments, {is expected} 
to switch from inertial to viscous. 
Note that the {crossover} 
air-layer thickness derived from such a balance, 
approximately $106 \mu$m, is still 
around 40 times larger than the air-layer thickness for which the cushioning-problem is initiated by the non-dimensionlisation of \citet{rosshicks2019} and \citet{moore2021introducing}.}\\




\section{Conclusion}
We report the results from an experimental study covering wedge and cone impacts into a bath of water. The impacts were done {over} a large range of velocities, {which were precisely controlled} using a linear motor {and kept constant during impact.} Pressure transducers with a $5.5$ mm wide sensing diaphragm were installed at two locations along opposite sides on a wedge and a cone{, both of which had a deadrise angle of 10$\degree$}. After the wedge's keel makes its first contact with the water, a {slight} under-pressure is registered prior to the peak pressure being attained at {that} location. Previous works suggested this to occur due to jet flow along the impactor surface. We show using {high {accuracy} measurements that these under-pressures are also rescalable by} {inertial} {time} and {length} scales.

{The entire pressure time series} are compared to Logvinovich \citep{logvinovich1969hydrodynamics,korobkin2005modified}, generalised Wagner \citep{korobkin_2004} and {composite} \citep{zhao_faltinsen_1993} models. The computed results from the models are spatially averaged over the sensing area of the transducers. {The comparisons offer important insight into predictive abilities of the (slightly) different approaches to avoiding the spatial singularity inherent in the original Wagner model. In general we are inclined to favour the \citet{zhao_faltinsen_1993} model as coming closest to the {measured space-averaged pressure peaks}, even if OLM, MLM and GWM predict the highest point-pressures (table \ref{tab:cpmaxcoeffs}}). {In the space-averaged calculations from OLM, MLM, GWM and \citet{zhao_faltinsen_1993} models, we find excellent reproducibility of how the peak pressure coefficients become larger along the length of the body.}

The theoretical treatment for wedge slamming is done assuming 2D flows resulting from impact. The treatment's extension for slamming of cones assumes axisymmetric flow in the outer domain, and 2D flow in the inner and jet domains. {Although theoretically plausible, t}he latter approximation has never been justified {directly in experiment. Here, we} do so by comparing the {predicted} ratio of peak pressures between a cone and wedge from theory $(64/\pi^4)$, to the same {ratio} measured from experiments. The agreement between the theory and experiments is found to be excellent. This allows us to conclude that the usual approach of treating the inner flow as two-dimensional is {sufficient} as far as the pressure distribution due to impact is concerned.

{The final part of the paper concerns {the} air cushioning prior to impact{, namely the downwards deflection of} the water surface, {which is mediated by the air phase} being squeezed out by the approaching impactor. In our lab scale setup, the {measured} final depth by which the target surface is depressed by the air cushioning layer is of the order $10^{-3}$ times the impactor diameter{, a result which is corroborated by the boundary integral simulations we performed}. An air layer will only be fully entrapped when the gradient of the deformed liquid surface is larger than that of the approaching finite-deadrise object. {For our experiments this condition is met only at very small deadrise angles of $\lessapprox 1\degree$ {(see supplementary material)}, but in a larger scale setting, compressibility effects or surface instabilities may make this condition easier to attain. Finally, it is good to note that for the slamming impacts studied in this work, the measured surface deformation due to inertial effects in the gas phase sets in long before viscous effects may become important. In the latter case, a micrometric air bubble may be entrapped as discussed by \citet{bouwhuisprl, bouwhuisinitial} and \citet{rosshicks2019}.}}

\section*{Acknowledgements}

We acknowledge financial support from SLING (project number P14-10.1), which is partly financed by the Netherlands Organisation for Scientific Research (NWO).

\section*{Declaration of Interests}
The authors report no conflict of interest.

\section*{Author ORCID}
\noindent U. Jain, \href{https://orcid.org/0000-0002-1014-7861}{https://orcid.org/0000-0002-1014-7861}\\D. van der Meer, \href{https://orcid.org/0000-0003-4420-9714}{https://orcid.org/0000-0003-4420-9714}

\appendix

\section{Inferring acceleration from placebo transducers}\label{sec:wedgesappendix}

\begin{figure}
    \centering
    \includegraphics[width=.99\linewidth]{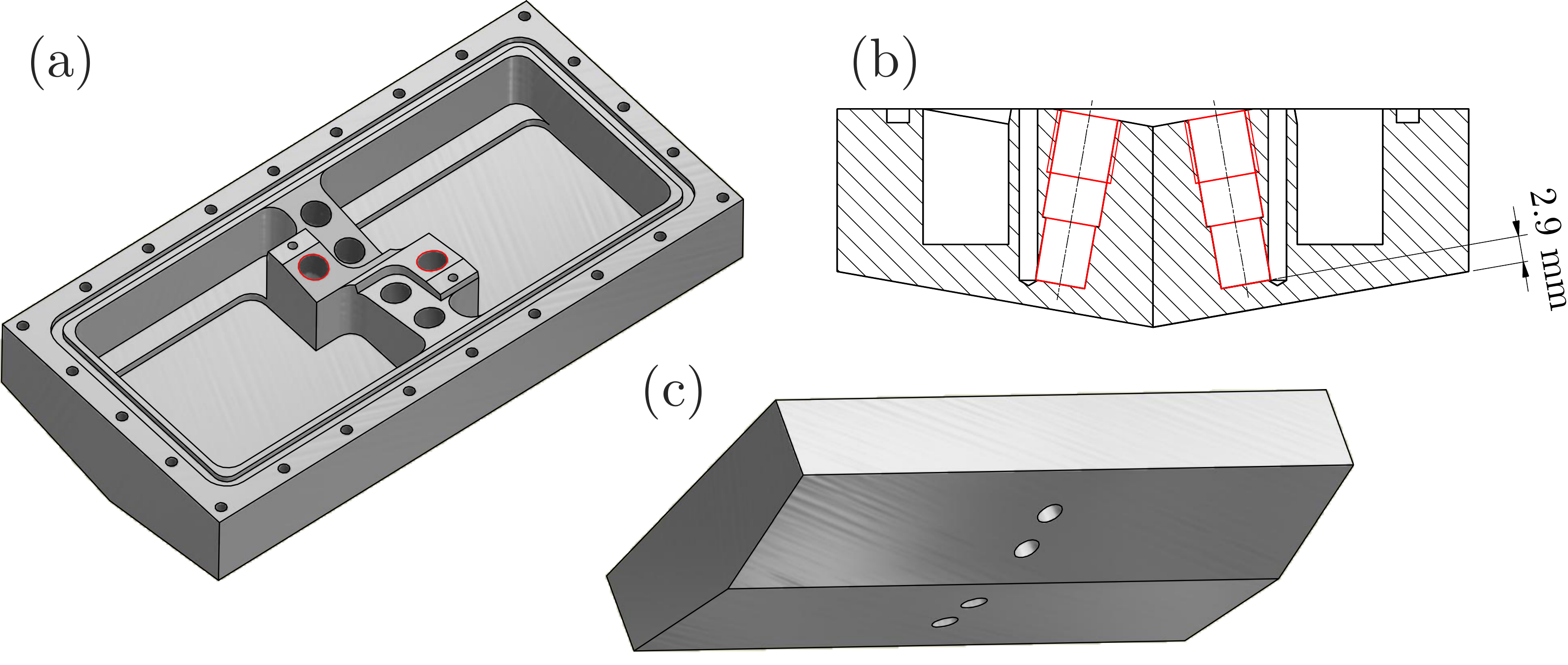}
    \caption{The overall design for the wedge used is shown. The sensor locations marked with red outline in panels (a) and (b) are where the blind sensors were installed. They were installed such that the sensing area was not flush mounted with the impacting surface. The impacting surface is shown in panel (c). }
    \label{fig:placebosensors}
\end{figure}

When performing impact tests with transducers installed on a moving part, there is always a suspicion that a {component of} the pressures {signals} may have originated due to the moving part's acceleration. We checked for this by installing blind `placebo' sensors on both the cone, and the wedge. These sensors are mounted such that their sensing area is not exposed to the liquid at the wedge's (or cone's) impacting surface. The wedge design is shown in figure \ref{fig:placebosensors}. Only the placebo sensors are shown; the other sensors were mounted at the same locations as the cone (as shown in figure \ref{fig:conedesign}).

\begin{figure}
    \centering
    \includegraphics[width=.85\linewidth]{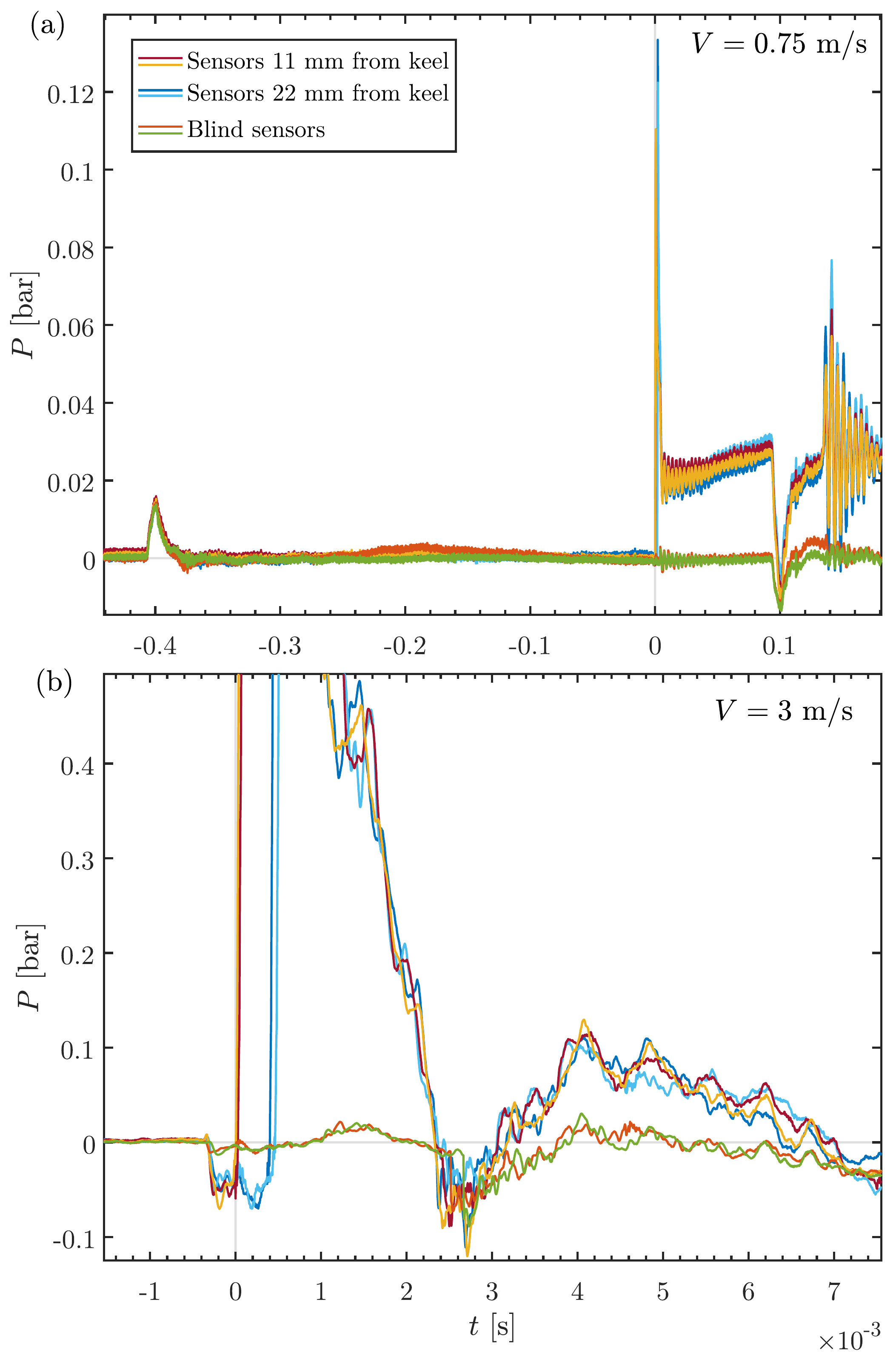}
    \caption{(a) Pressure measurements from all sensors at $V=0.75$ m/s and {(b) at $V=3$ m/s. In panel (a) we show both the start and end of the impactor's motion. In (b) we only show the region closer to impact to show the differences between the exposed and blind sensors.}}
    \label{fig:timeserieswithplacebos}
\end{figure}

A set of measurements from a wedge at $V = 0.75$ m/s are shown in figure \ref{fig:timeserieswithplacebos}(a). At $t = -0.4$ s it can be seen that the reading on all the sensors suddenly increases from the earlier base value of zero bar. This is seen to occur when the wedge's motion is started from rest outside of the target liquid. The time series are centred about $t=0$ ms when the first set of sensors (11 mm from the keel) register their peak pressures. At this stage it can be seen that only the readings on sensors flush mounted at the impacting surface suddenly rise. At the same time the readings on the blind sensors vary negligibly from the base reference of zero bar. After the impact peak, at approximately $t = 0.1$ s, the readings on all the sensors undergo a sudden drop. This corresponds to a sudden retardation of the wedge after impact. Interestingly, both the retardation and the acceleration of the wedge cause a significant change in the pressure readings on both the exposed and blind sensors. {It is also worth noting that both the acceleration and deceleration, at the beginning and end of the wedge's motion, register the same pressures on both exposed and blind sensors.} In contrast at impact, only the exposed sensors register a significant change in pressure. On one hand this suggests that the through the impact process, a constant velocity of the impactor was reasonably well maintained. More clearly, the difference in the behaviour of pressures on the blind and flush-mounted sensors allows us to conclude that the measurements reported throughout the manuscript are at the most negligibly affected by the impactor's acceleration. {The same measures at a much higher velocity 3 m/s are shown in figure \ref{fig:timeserieswithplacebos}(b). The timescales at this velocity are much shorter, so the results are zoomed-in into the times around impact--deceleration. At this higher $V$ the blind sensors show some added-mass response at impact ($t=0$) by going lower than the reference pressure. In this figure the moment of wedge's stopping is at approx 2.65 ms. Same as panel (a), both the exposed and blind sensors register the sharp deceleration at the moment of wedge becoming stationary.}

{\section{{Impact force on cones} }}
\begin{figure}
    \centering
    \includegraphics[width=.99\linewidth]{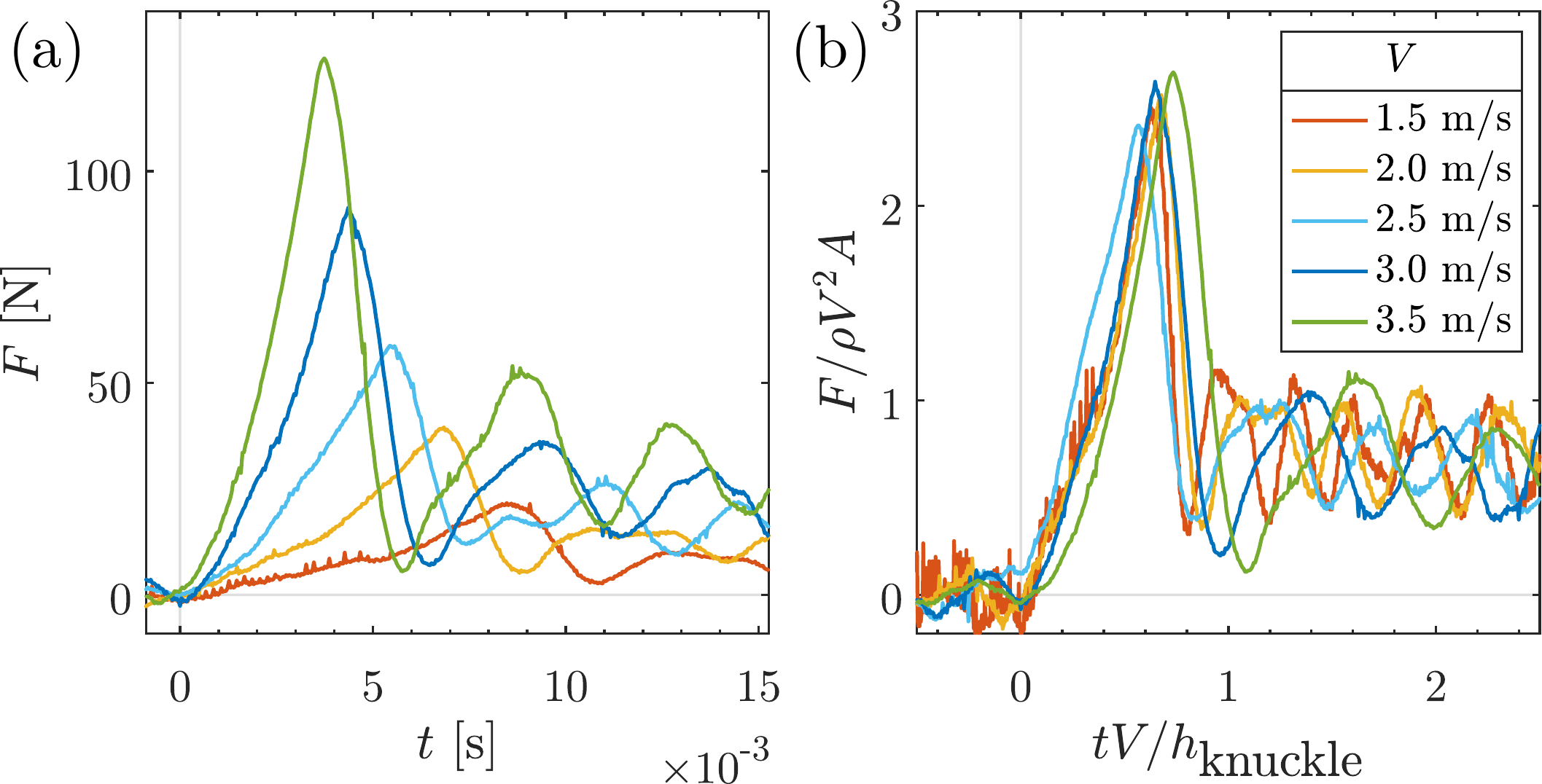}
    \caption{{{(a) Time series of the force $F$ during impact of a cone with a deadrise angle of $30\degree$, measured for five different impact velocities. (b) The same data as in (a) but now nondimensionalized with inertial force ($\rho V^2 A$) and time ($h_\text{knuckle}/V$) scales, leading to a fair collapse of the data.}  }}
    \label{fig:forcetimeseries}
\end{figure}

\begin{figure}
    \centering
    \includegraphics[width=.99\linewidth]{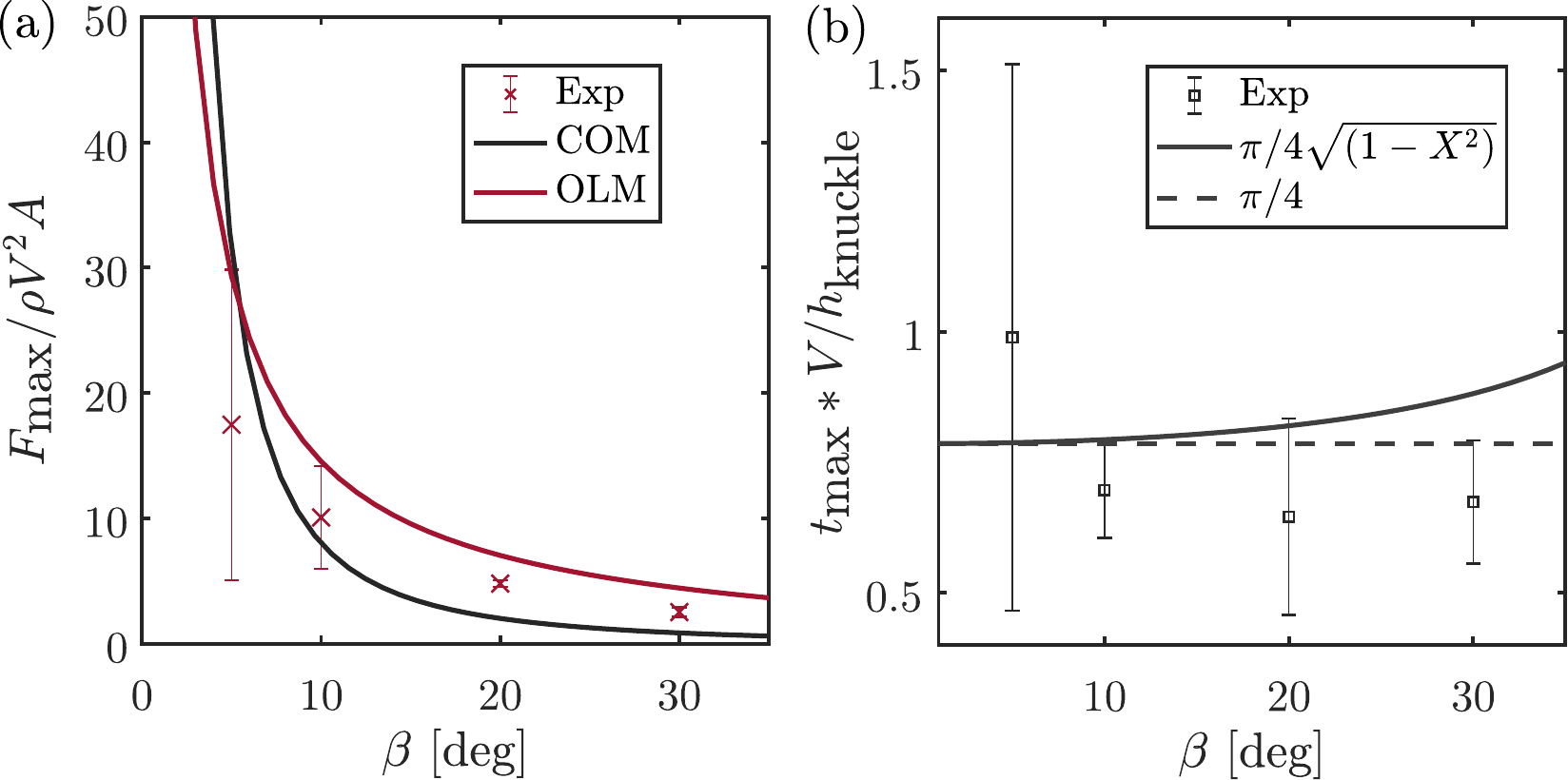}
    \caption{{{(a) The measured peak force coefficients $F_\text{max}/(\rho V^2 A)$ are plotted against the deadrise angle $\beta$ of the cone and compared to the estimated peak force coefficients obtained from the original Logvinovich (OLM) and composite (COM, \citet{zhao_faltinsen_1993}) models.} (b) The {experimental non-dimensionalised times $t^*_\text{Fmax} = t_\text{Fmax} V / h_\text{knuckle}$ at which the peak force $F_\text{max}$ occurs,} is compared to estimations {obtained from $c(t) = D/(2\cos\beta)$ (solid black line) and $a(t) = D/(2\cos\beta)$ respectively (dashed line).}}}
    \label{fig:forcepeaksandtmax}
\end{figure}

{{In addition to the pressure measurements discussed in the main text, total impact forces were measured on cones with $\beta =$ 5, 10, 20 and 30$\degree$ deadrise angles using FUTEK LSM300 analog strain gauge load cells. {Depending on the expected maximum force, one of two load cells with a different} maximum loading capacity of 200 or 500 lbs was used, with measured system} natural frequencies {of $1.7$ kHz and $2,5$ kHz respectively}. The load cells were calibrated using known weights, and the data was read out done using an analog current output amplifier (FUTEK IAA200) at an acquisition rate of 25 kHz.} 

{The result of these measurements for a $30\degree$ deadrise-angle cone are plotted in figure \ref{fig:forcetimeseries}(a) for five different impact velocities $V$. From the time series we observe that the force $F$ is first increasing steeply, reaches a maximum approximately when the jet base of the flow has reached the knuckle, and then suddenly decreases. After this, the force starts to oscillate with the resonant frequency of the system. Rescaling both the force $F$ and time $t$ with the inertial scales $\rho V^2 A$ (with $A = \pi D^2/4$ the frontal area of the cone) and $h_\text{knuckle}/V = \tan\beta D/V$, we find a fair data collapse in the region corresponding to the impact, which of course is lost during the oscillations since they do not follow the same inertial scaling as the impact.}

{Subsequently, we compare the measured nondimsionalized peak forces for different cones to theoretical models.} {The force on a cone can be estimated by {integrating the equivalent of the wedge pressure expressions for OLM, MLM and GWM (\eqref{eqn:OLM}--\eqref{eqn:GWM}) for the cone case}, or the cone version of the composite model (COM) \eqref{eqn:compositepressure2}{, over the surface area of the cone. Since of the former three models, only the wetting correction for} original Logvinovich model (OLM) {was available for cone, we restrict ourselves to OLM.} Integrating \eqref{eqn:OLM} for a cone yields 
\begin{align} \label{eqn:logvinovichconeforce}
    F(t) &= \int_{0}^{a(t)}  x dx  \int_{0}^{2\pi} d\theta \; P(x,t) \nonumber\\
    &= - \frac{\rho V^2}{2}\pi c  \left( \frac{ 2\pi \sqrt{c^2 - a^2}}{\tan \beta}  - c \log \left(c^2 - a^2 \right)  - \frac{ 2 \pi c}{\tan \beta}  + c \log\left( c^2\right)  \right) \,,
\end{align}}{where we note that the factor $\cos^{-1}\beta$ appearing in the surface element [$(x(\cos^{-1}\beta dx)d\theta$] is cancelled by the fact that we need to multiply with $\cos\beta$ to single out only the vertical component of the force.} {Here, as in the main text (expression \eqref{eqn:X_OLM}) $a(t)=\xi c(t)$, with $\xi = \sqrt{1-X^2}$, but $X$ needs to be replaced by the corresponding expression that is valid for the cone case: $X = \pi \tan \beta /4$. A similar expression can be obtained using the composite model.} 

{The force calculated in expression \eqref{eqn:logvinovichconeforce} is rising continuously and monotonically in time, since no information of the finite size of the cone has been included. To do so, we infer that the maximum force $F_\text{max}$ on the cone is reached when the jet base has reached the knuckle, i.e., when $c(t) = x_\text{knuckle} = D/2$.} {We compute $F_\text{max}$ as a function of the deadrise angle $\beta$ by inserting $c(t) = D/2$ in equation \eqref{eqn:logvinovichconeforce} and compare the result with the experimentally determined force, after nondimensionalising both with the inertial force scale $\rho V^2 A$ (with $A = \pi D^2/4$), in figure \ref{fig:forcepeaksandtmax}(a).}
{Similarly, the time when peak force is attained, $t_{\text{Fmax}}$, can be estimated by assuming that {at that time, $c(t)=D/2$, i.e.,} the liquid completely wets the cone surface. {Rearranging this condition and nondimensionalising $t_{\text{Fmax}}$ using $h_{\text{knuckle}}/V$} gives $t^*_{\text{Fmax}} = {\pi /4}$. Another estimation may be obtained by saying $F_{\text{max}}$ is attained when $a(t)=D/2$. Rearranging this equality and using $a=\sqrt{1-X^2}\;c$, one finds $t^*_\text{Fmax} = { \pi  / (4\sqrt{1-X^2})}$. These estimations are compared to the measured $t^*_{\text{Fmax}}$ in figure \ref{fig:forcepeaksandtmax}(b) finding reasonable agreement.}\\


{\section{Space-averaging pressures from models}\label{sec:spaceavappendix}}

\begin{figure}
    \centering
    \includegraphics[width=.9\linewidth]{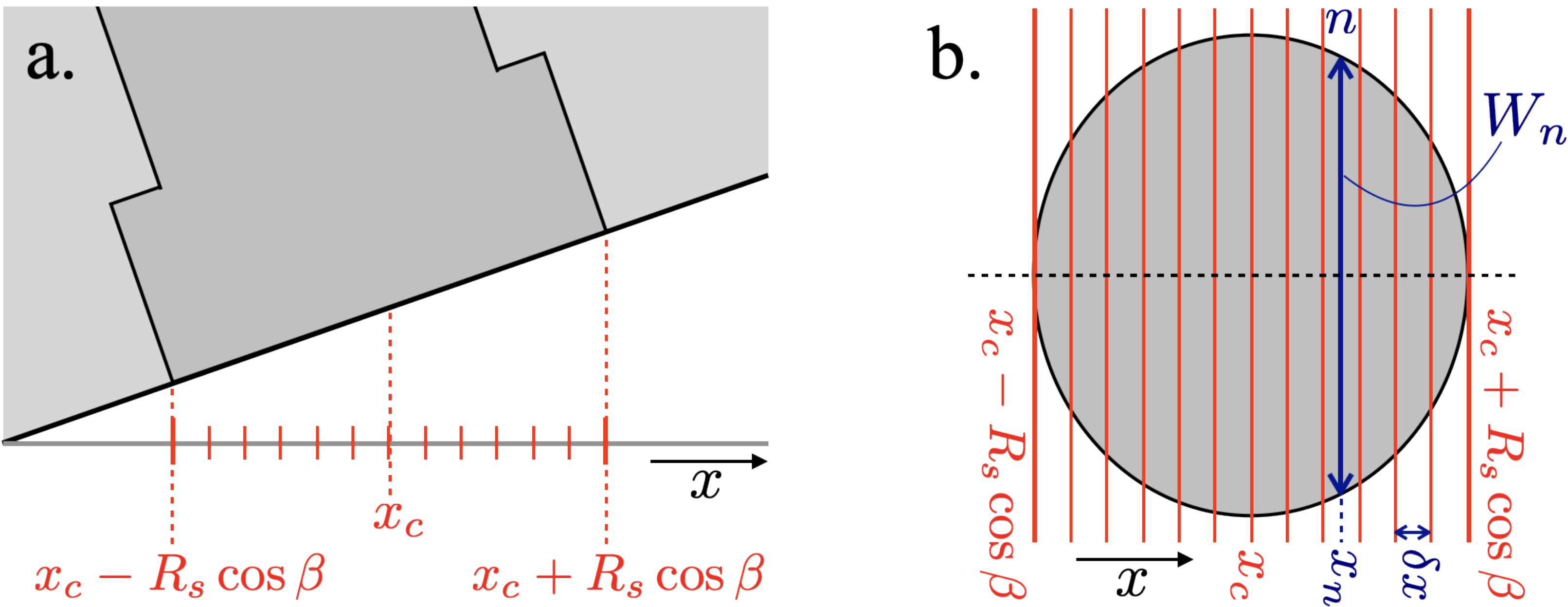}
    \caption{{Sketch of the sensor's projection, explaining the variables in equations \eqref{eqn:averaging1} and \eqref{eqn:averaging2} to do a space-average of the point-pressures from models \eqref{eqn:compositepressure1} and \eqref{eqn:OLM}--\eqref{eqn:GWM}. }}
    \label{fig:projectionsketch}
\end{figure}
{To space average the point pressures computed from the various theoretical models, we discretize the sensor's surface area (or rather its projection onto the horizontal plane) into $N$ horizontal bins, that is, perpendicular to the flow direction on the wedge or cone{, as sketched in figure~\ref{fig:projectionsketch}.} Denoting the circular membrane radius of the sensor by $R_s$ and the $x$-coordinate of the center of the sensor as $x_c$, the sensor ranges from $(x_c-R_s\cos\beta)$ to $(x_c+R_s\cos\beta)$. Subdividing this range into $N$ bins of equal width $\delta x = 2R_s\cos\beta/N$, with centers $x_n$, we compute the point pressure $P(x_n,t)$ in each bin and sum all pressures with a weight $W_n$ that corresponds to the local width of the elliptic projection of the sensor at the bin at position $x_n$, that is
\begin{equation}\label{eqn:averaging1}
W_n = 2 \sqrt{R_s^2 - \left(\frac{x_n - x_c}{\cos\beta}\right)^2}.
\end{equation}} 
{Clearly, the total area of the sensing area is then simply proportional to} {the sum of the weights ($\sum\limits_{n=1}^{N} W_n$), such that the sensor-averaged pressure $P_\text{ave}(x_c,t)$ can be computed as 
\begin{equation}\label{eqn:averaging2}
   P_\text{ave}(x_c,t) =  \frac{\sum\limits_{n=1}^{N} P(x_n,t)  W_n}{\sum\limits_{n=1}^{N} W_n}\,.
\end{equation}}  


{\section{Inertial vs. viscous forcing of the interface prior to impact}\label{sec:inertialviscouscushioning}}

{The results presented in this study indicate that, for the explored parameter space and the time and length scales {relevant to cushioning before a slamming event}, the observed surface deformation can be attributed to inertial forcing, and viscous effects in the air phase may be neglected. In this appendix we show that this is consistent with theoretical results available in the literature.}

{\cite{bouwhuisinitial} compared the pressure on an initially flat liquid-air interface {when} a solid sphere of radius $R$ approach{es} with a constant velocity $V$, for which they compute the following approximate expressions for the inertial ($p_\textrm{in}$) and viscous ($p_\textrm{vis}$) contributions
\begin{eqnarray}    
p_\textrm{in} &=& \rho_gV^2\left[\left(\frac{R}{\zeta}\right)^3\!\!F_2(u)+\tfrac{9}{2}\left(\frac{R}{\zeta}\right)^6\!\!F_3(u)\right]\quad\textrm{with:}\,\,\,u=\frac{r}{\zeta}\,,\,\,\zeta = R + h_0\,,\label{eqA:p_in}\\ 
p_\textrm{vis} &=& \frac{3\eta_gV}{R}\left(\frac{R}{L}\right)^4\!\!F_1(\tilde{u}) \quad\textrm{with:}\,\,\,\tilde{u}=\frac{r}{L}\,,\,\,L = \sqrt{Rh_0}\,,\label{eqA:p_vis}
\end{eqnarray}
where $h_0 =  V(t_i - t)$ is the vertical distance between the sphere and the (undeformed) liquid interface, $r$ the radial distance to the symmetry axis of the impact, and $\eta_g$ and $\rho_g$ are the dynamic viscosity and density of the air phase. Note that $t$ denotes time and $t_i$ is the expected time of impact on the (undisturbed) interface. The full expressions of the functions $F_1$, $F_2$, and $F_3$ are available in \citet{bouwhuisinitial}, but for the purpose of this appendix it is sufficient to know that they rapidly converge to zero for large values of their argument ($u,\tilde{u} \gtrsim1$) and to provide their values in $r = 0$, for which in both cases the pressure is maximum: $F_1(0) = 1$, $F_2(0) = 2$, and $F_3(0) = 0$.} 

{It is good to note that technically \eqref{eqA:p_in} is expected to be valid only for large $h_0$ and  \eqref{eqA:p_vis} for small $h_0$, such that for quantitatively accurate predictions, corrections should be applied to both expressions in the region $h_0 \approx R$ but that this is expected to have little impact on the qualitative arguments provided below. In fact, in \citet{bouwhuisinitial} an alternative small gap expression for the inertial pressure is derived for small $h_0$, which is found to scale similarly as $p_\textrm{vis}$, but is shown to be vanishingly small.}   

{The first observation that one can make from equations~\eqref{eqA:p_in} and \eqref{eqA:p_vis} is that, in the case of the inertial pressure $p_\textrm{in}$, for small $h_0$ the typical length scale $\zeta$ of the profile is determined by the size of the sphere: $\zeta \sim R$, whereas the typical length scale $L$ of the viscous pressure profile $p_\textrm{vis}$ is determined by the vanishing distance to the interface: $L \sim \sqrt{h_0}$. This implies that the surface deformation caused by air inertia is always on the length scale of the impacting sphere, whereas the deformation caused by air viscosity becomes smaller and smaller as the moment of impact approaches. Physically, this very local deformation in the latter case originates from the fact that lubrication pressures increase strongly with decreasing gap width $h_0$ \citep{rosshicks2019,hicks_ermanyuk_gavrilov_purvis_2012, smith_li_wu_2003}}. 

{Computing $p_\textrm{in}$ and $p_\textrm{vis}$ at the symmetry axis of the impact leads to 
\begin{eqnarray}    
p_\textrm{in}(0,t) &=& 2\rho_gV^2\left(\frac{R}{R + h_0}\right)^3 \,,  \label{eqA:p_in_ax}\\ 
p_\textrm{vis}(0,t) &=& \frac{3\eta_gV}{R}\left(\frac{R}{h_0}\right)^2\,.\label{eqA:p_vis_ax}
\end{eqnarray}
From these expressions it is clear that at large $h_0$ the inertial pressure is dominant, whereas for small $h_0$ the viscous contribution dominates. From equating the above two expressions one may (for $h_0 \ll R$) approximate the crossover distance $h_{0,c}$ as  
\begin{equation}
\frac{h_{0,c}}{R} \approx \sqrt{\frac{3}{2 \textit{Re}_g}}\quad\textrm{with:} \,\,\textit{Re}_g = \frac{\rho_g V R}{\mu_g}\,.\label{eqA:crossover}
\end{equation} 
This approximation is justified since the Reynolds number $\textit{Re}_g$ is typically large, and consequently the distance $h_0$ at which viscous effects become dominant is small.}

{We may estimate the deformation caused by inertia and viscous effects by time-integrating the pressure contributions \eqref{eqA:p_in_ax} and \eqref{eqA:p_vis_ax}, which leads to estimates for the inertial ($\Pi_\textrm{in}$) and viscous ($\Pi_\textrm{vis}$) contributions to the pressure impulse imparted to the water surface
\begin{eqnarray}    
\Pi_\textrm{in} &=& \int_{t=-\infty}^{h_{0,c}/V} p_\textrm{in}(0,t)dt = \frac{\rho_gVR^3}{(R+h_{0,c})^2} \approx  \rho_gVR = \textrm{Re}_g\,\eta_g\,,  \label{eqA:impulse_in_ax}\\ 
\Pi_\textrm{vis} &=& \int_{h_{0,c}/V}^t p_\textrm{vis}(0,t)dt = 3\eta_g \left[\frac{R}{h_0} - \sqrt{\tfrac{2}{3}\textit{Re}_g}\right] \sim \sqrt{\textrm{Re}_g}\,\eta_g \,,\label{eqA:impulse_vis_ax}
\end{eqnarray}
where the last similarity is intended to express what happens just after viscosity has become appreciable noting that the two terms balance exactly for $h_0 = h_{0,c}$. In conclusion, the deformation caused by air inertia is already appreciable on the length scale of the impacting sphere ($\sim R$), when air viscosity becomes important and causes further interface deformation on the length scale $L = \sqrt{h_{0,c}R} \approx R (2\textit{Re}_g/3)^{-1/4}$ and smaller. Clearly, this viscous deformation will therefore emerge at a later point in time and will have a smaller range.     }

{For a cone (or wedge) with not too large deadrise angle (as used in our experiments), what will change in the above picture is that a surface deformation due to inertia will emerge on the length scale of the entire cone ($R$), but the viscous contribution will be determined by the curvature of the keel ($R^*$), which is considerably smaller than $R$. Replacing $R$ with $R^*$ in expression \eqref{eqA:p_vis_ax} and recomputing the crossover distance leads to 
\begin{equation}
\frac{h^*_{0,c}}{R} \approx \sqrt{\frac{R^*}{R}}\sqrt{\frac{3}{2 \textit{Re}_g}}\,.\label{eqA:crossovercone}
\end{equation} 
Inserting typical parameters ($R = 3.5$ cm, $R^* = 0.5$ mm, $V = 1.0$ m/s, $\rho_g = 1.2$ kg/m$^3$, $\eta_g = 18$ \textmu Pa$\cdot$s) we find that $h^*_{0,c} \approx 106$ \textmu m. This implies that viscous deformations only become appreciable on a length scale smaller than $\sqrt{Rh^*_{0,c}} \approx 2.3$ mm and at times closer than $\Delta t = h^*_{0,c}/V \approx 106$ \textmu s before impact. Clearly, these length and time scales are difficult to resolve experimentally.}

{In conclusion, whether surface deformations originating from inertial or viscous forcing in the air phase are most important depends strongly on the question one wants to answer. If one focuses on local effects, such as determining the maximal pressure occurring at the very first contact or at estimating the size of the microbubble entrapped upon impact (as in \cite{bouwhuisinitial}), viscous effects are crucial. For the overall deformation of the free surface however, inertial effects dominate and viscosity may safely be neglected. }


\bibliographystyle{jfm}
\bibliography{jfm-instructions}

\end{document}